\documentclass[10pt,conference]{IEEEtran}

\usepackage{times}
\usepackage{amsmath}
\usepackage{amssymb}
\usepackage{epsfig}

\newcommand{\eqnlabel}[1]{\label{eqn:#1}}
\newcommand{\eqnref}[1]{(\ref{eqn:#1})}

 \providecommand{\Yv}{Y^N}
 \providecommand{\uv}{u^N}
\providecommand{\xv}{x^N} \providecommand{\yv}{y^N}

\providecommand{\Nc}{{\cal N}}
\providecommand{\Xc}{{\cal X}}
\providecommand{\Yc}{{\cal Y}}

\newtheorem{MyTheorem}{Theorem}
\newcommand{\thmlabel}[1]{\label{thm:#1}}
\newcommand{\thmref}[1]{\ref{thm:#1}}

\newtheorem{MyLemma}{Lemma}
\newcommand{\lemmalabel}[1]{\label{thm:#1}}
\newcommand{\lemmaref}[1]{\ref{thm:#1}}

\newtheorem{MyCorollary}{Corollary}
\newcommand{\corlabel}[1]{\label{thm:#1}}
\newcommand{\corref}[1]{\ref{thm:#1}}

\newtheorem{MyRemark}{Remark}

\title{On the Capacity of Interference Channels with \\
 One Cooperating Transmitter}
\begin{document}

\author{
\authorblockN{Ivana Mari\'c}
\authorblockA{Stanford University\\
Stanford, CA \\
ivanam@wsl.stanford.edu} \and
\authorblockN{Andrea Goldsmith}
\authorblockA{Stanford University  \\
Stanford, CA \\
andrea@wsl.stanford.edu } \and
\authorblockN{Gerhard Kramer}
\authorblockA{Bell Labs, Alcatel-Lucent \\
Murray Hill, NJ \\
gkr@bell-labs.com} \and
\authorblockN{Shlomo Shamai (Shitz)}
\authorblockA{Technion \\
Haifa, Israel \\
sshlomo@ee.technion.ac.il} }

\maketitle \thispagestyle{empty} \pagestyle{empty} \footnotetext[1]{
The work by I.~Mari\'c and A. Goldsmith was supported in part from
the DARPA ITMANET program under grant 1105741-1-TFIND, Stanford's
Clean Slate Design for the Internet Program and the ARO under MURI
award W911NF-05-1-0246. The work of G.~Kramer was partially
supported by the Board of Trustees of the University of Illinois
Subaward No.~04-217 under NSF Grant~No.~{\scriptsize CCR-0325673}.
The work of S. Shamai was supported
by the EU 7th framework program: NEWCOM++.}
%%%%%%%%%%%%%%%%%%
\begin{abstract}
  Inner and outer bounds are established on the capacity region
   of two-sender, two-receiver interference channels where one
   transmitter knows both messages. The transmitter with extra
   knowledge is referred to as being cognitive. The inner bound is based on strategies that generalize
   prior work, and include rate-splitting, Gel'fand-Pinsker coding and cooperative transmission.  A general outer
   bound is based on the Nair-El Gamal outer bound for broadcast
   channels. A simpler bound is presented for the case in which one of the decoders can decode both messages. The bounds are evaluated and compared for Gaussian channels.
\end{abstract}
%%%%%%%%%%%%%%%%%%%%
%%%%%%%%%%%%%%%%%%%%%%%%%%%%%%%%%%%%%%%%
\section{Introduction and Related Work}
%%%%%%%%%%%%%%%%%%%%%%%%%%%%%%%%%%%%%%%%
Two-sender, two-receiver channel models allow for various forms of
transmitter cooperation. When senders
are unaware of each other's messages, we have the interference
channel \cite{Sato77,Carleial78}.
In wireless networks, the broadcast nature of the wireless medium allows nodes to overhear transmissions and possibly decode parts of other users' messages.  An encoder that has such knowledge can use it to improve its own rate and the
other user's rate. The level of cooperation and performance improvement will
depend on the amount of information the encoders share.
In the interference channel,  rate gains from the transmitter cooperation were demonstrated in~\cite{DevroyeSubmitted}.

\begin{figure}[t]
\center\epsfig{figure=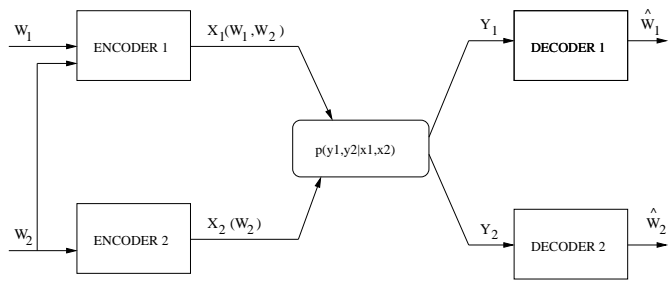,width=3.5in} \caption[Interference
channel with cooperating encoder.]{Interference channel with
cooperating encoder.}
      \label{f:Figure}
\end{figure}

%%%%%%%%%%%%%%%%%%%
% COGNITIVE RADIO
%%%%%%%%%%%%%%%%%%%
Channel models with cooperating nodes are of interest also for networks
with cognitive users. Cognitive radio \cite{Mitola} technology is
aimed at developing smart radios that are both aware of and adaptive
to the environment. Such radios can  efficiently sense the spectrum,
decode information from detected signals
 and use that knowledge to improve the system performance. This
technology motivates information-theoretic models that try to
capture the cognitive radio characteristics.
In that vein, this paper considers a two-sender, two-receiver channel
model in which, somewhat
idealistically, we assume  that cognitive capabilities allow one user to know the full message of the other encoder, as shown in Fig.~\ref{f:Figure}. Existing encoding schemes can  bring different rate gains that depend on the channel characteristics and topology, making it challenging to determine the capacity even for special cases. This paper is a step along this path.
It would further be interesting to extend the existing results to large networks with cooperating encoders.

%offers insights into how the  gains are achieved and into ways to extend these approaches to large %networks with cooperating encoders.
%%%%%%%%%%%%%%%%%%%%%%%%%%%%%%%%%%%%%%%%%%%%
%%%%%%%%%%%%%%%%%%%%
%%%%%%%%%%%%%%%%%%%%%%%%%%%%%%%%%%%%%%%%%%%%%%%%%%
% RELATED WORK
%%%%%%%%%%%%%%%%%%%%%%%%%%%%%%%%%%%%%%%%%%%%%%%%%%
Investigating the capacity region of the channel model we consider is the focus of much recent work.
In particular, the interference channel with one cooperating encoder was dubbed the
{\it cognitive radio channel} and achievable rates were presented in
\cite{DevroyeSubmitted, DevroyeCommMag}.
A general encoding scheme was also proposed more recently in
\cite{JiangXin2007}.
  The capacity region for the Gaussian case of weak
interference was determined in~\cite{WuVishwanath06} and
\cite{JovicicViswanath2006}.
The results of~\cite{WuVishwanath06,JovicicViswanath2006} were extended to the Gaussian MIMO cognitive radio network and shown to achieve the sum-capacity in \cite{SridharanVishwanath2007}.
Related work can also be found in \cite{Tuninetti2007, CaoChenZhang2007}.
However, the conclusions of \cite{SridharanVishwanath2007} do not immediately apply to the single-antenna cognitive radio channel.
%%%%%%%%%%%%%%%%%%%%%%
% RESULTS IN THIS PAPER
%%%%%%%%%%%%%%%%%%%%%%%
In this paper, we  present a scheme that generalizes those in
\cite{WuVishwanath06}-\cite{MGKSITA2007}.
%For the Gaussian channel, the conclusions of \cite{SridharanVishwanath2007} immediately imply that the %presented scheme achieves the sum-capacity.
The scheme is similar to
the one in \cite{JiangXin2007}: as in \cite{JiangXin2007} and
\cite{DevroyeSubmitted}, an encoder uses {\it rate-splitting}
 \cite{Carleial78} to enable the other receiver to decode part of the
 interference; the cognitive transmitter
 cooperates in sending the other
 user's message to its intended receivers and uses
Gel'fand-Pinsker (GP) binning  \cite{GelfandPinsker80} to reduce interference to its own receiver.
The key difference of our contribution to the prior work is in the way the binning is
performed.
An overview of the encoding scheme is given in the next section.
The encoding scheme is derived in  Section~\ref{sect:achievablerates}, compared to other results and adapted for Gaussian channel in Section \ref{sect:Gaussianchannel}.

%%%%%%%%%%%%%%%%%%%%%%%%%%%%%%%%%%%%%%%
% OUTER BOUNDS
%%%%%%%%%%%%%%%%%%%%%%%%%%%%%%%%%%%%%%%
In Section~\ref{sect:outerbounds}, we
present two outer bounds for the interference channel with one cooperating encoder. The first bound is based on the Nair-El Gamal broadcast outer
bound, \cite{NairElGamal2007}. It has the same mutual information expression as the one in \cite{NairElGamal2007}, the only difference is in the input distribution over which the optimization is performed. The bound thus reflects the resemblance of the considered channel to the broadcast channel (BC), and the difference given by the fact that encoder $2$ has only partial knowledge of messages sent in the channel. We then present an outer bound for the strong interference case that is of the same form as the one in \cite[Sect.V]{MYKsubmitted}, and compare it to the achievable rate region in Gaussian channels.
 Results also demonstrate an improvement compared to the general scheme of \cite{JiangXin2007}.

%%%%%%%%%%%%%%%%%%%%%%%%%%%%%%%%%%%%%%%%
\section{Overview of the Encoding Strategy}
%%%%%%%%%%%%%%%%%%%%%%%%%%%%%%%%%%%%%%%%
%%%%%%%%%%%%%%%%%%%%%%
% 1. Rate-splitting
%%%%%%%%%%%%%%%%%%%%%%
The considered channel model has elements of both the interference channel (IC) and the broadcast channel (BC). Encoding techniques developed for either of them are therefore potentially useful.
If the message $W_2$ of encoder $2$ was not known at the cognitive encoder, the considered channel would reduce to the interference channel (IC). The best achievable rate region for the IC, \cite{HanKobayashi81}, is achieved by rate-splitting \cite{Carleial78}: each encoder divides its message into two parts and encodes each of them with a separate codebook. This allows receivers to decode one of the two sub-messages of the other user's and cancel a part of the interference that it would otherwise create. Rate-splitting in the cognitive radio channel model was applied in \cite{DevroyeSubmitted, JiangXin2007}. In this paper, rate-splitting is performed at the cognitive encoder.
%%%%%%%%%%%%%%%%%%%%%%%%%%%%%%%%%%%%%%%%%%%%%%%%%%%%%%%

Additional knowledge  allows the cognitive encoder to employ a number of techniques in addition to rate-splitting. In particular, to improve the rate for the noncognitive communicating pair, the cognitive encoder can {\it cooperate} by encoding $W_2$ to help convey it to the other decoder.
On the other hand, any signal carrying information about $W_2$ creates interference to the cognitive encoder's receiver. This interference is known at the cognitive transmitter and the precoding technique, i.e. Gel'fand-Pinsker binning \cite{GelfandPinsker80} and, specifically, dirty-paper coding (DPC) \cite{Costa83} in Gaussian channels, can be employed.
%The situation is depicted in Fig.~\ref{X2asinterference} where $S$ denotes the interference created by the codebook or rate $R_s$ interfering with the communication of message $W$ at rate $R$.
In fact, GP binning is crucial for the cognitive radio channel: together with cooperation, it leads to capacity in certain scenarios, \cite{WuVishwanath06,JovicicViswanath2006,SridharanVishwanath2007}.
It is not surprising that  DPC  brings gains in the Gaussian cognitive radio channel:
if the non-cognitive encoder is silent, we have the broadcast channel from the cognitive encoder to two receivers, for which dirty-paper coding is the optimal strategy \cite{WSS06, Vishwanath03}.

In general, however, there are two differences at the cognitive encoder  from the classical GP setting.
First, the interference carries useful information for receiver $2$. Second, the interference is a {\it codebook} of some rate and can thus have lower entropy than in the GP setting. As we will see in Sect.~\ref{ss:binningcodebook}, the latter can be exploited to achieve a higher rate.

We note that due to  rate-splitting, there is a common part of $W_1$ decoded at the both receivers and precoded against interference. Since the signal carrying this common message experiences {\it different} interference at the two receivers, we use the ideas of \cite{Mitran2006} and \cite{KhistiErez06} that respectively extend \cite{GelfandPinsker80} and \cite{Costa83} to channels with different states non-causally known to the encoder. In the Gaussian channel, dirty paper coding is generalized to  {\it carbon-copying} onto dirty paper \cite{KhistiErez06} to adjust to the interference experienced at both  receivers.

%%%%%%%%%%%%%%%%%%%%%%%%%%%%%%%%%%%%%%%%%%%%%%%%%%%%%%
\subsection{Summary of Techniques and Special Cases}
%%%%%%%%%%%%%%%%%%%%%%%%%%%%%%%%%%%%%%%%%%%%%%%%%%%%%%
Although the interference channel with one cooperating encoder can easily be visualized as an extension of the classical IC, a number of techniques become potentially relevant due to additional knowledge of the cognitive encoder:
\begin{itemize}
\item Rate splitting at encoder $1$: Improves rate $R_2$ through interference cancelation at decoder $2$.
\item GP binning and binning against a codebook: Improves rate $R_1$ by precoding against interference. It also allows decoder $1$ to decode message $W_2$ (or part of it) when $R_2$ is small, as will be shown in Sect.~\ref{ss:binningcodebook}.
    \item Carbon-copying onto dirty paper: further improves the rate of the common message sent at the cognitive encoder
    \item Cooperation: Encoder $1$ contributes to rate $R_2$ by encoding $W_2$.
\end{itemize}
A general encoding scheme that brings these techniques together is described in Section~\ref{sect:achievablerates}. There will be number of special cases for which a  subset of techniques will suffice:
\begin{enumerate}
\item Strong interference: Both decoders can decode both messages with no rate penalty, so there is no need for either rate-splitting or binning. Superposition coding achieves capacity, \cite{MYKsubmitted}.
\item Cognitive encoder decodes both messages: Again, there is no need for binning. Rate-splitting and superposition coding achieve capacity, \cite{JiangXinGarg2007, LiangBaruchPoorShamaiVerdu2007}.
\item Weak interference at receiver $2$: There is no need for common part of message $W_1$ and hence for rate-splitting. Dirty paper coding and cooperation achieve capacity in Gaussian channel, \cite{WuVishwanath06,JovicicViswanath2006,SridharanVishwanath2007}.
\end{enumerate}
%%%%%%%%%%%%%%%%%%%%%%%%%%%%%%%%%%%%%%%%%%%%%%%%%%%%%%%%%%%%%%%%
\subsection{Rate Improvement due to Binning Against Codebook} \label{ss:binningcodebook}
%%%%%%%%%%%%%%%%%%%%%%%%%%%%%%%%%%%%%%%%%%%%%%%%%%%%%%%%%%%%%%%%
\begin{figure}[t]
\center\epsfig{figure=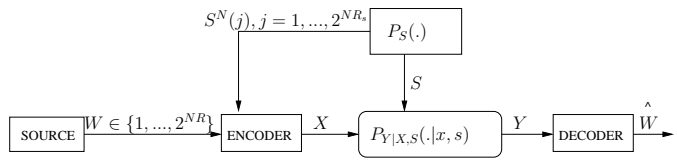,width=3.5in} \caption[Interference
channel with unidirectional cooperation.]{Communication from the cognitive transmitter to the corresponding receiver.}
      \label{X2asinterference}
\end{figure}
\begin{figure}[t]
\center\epsfig{figure=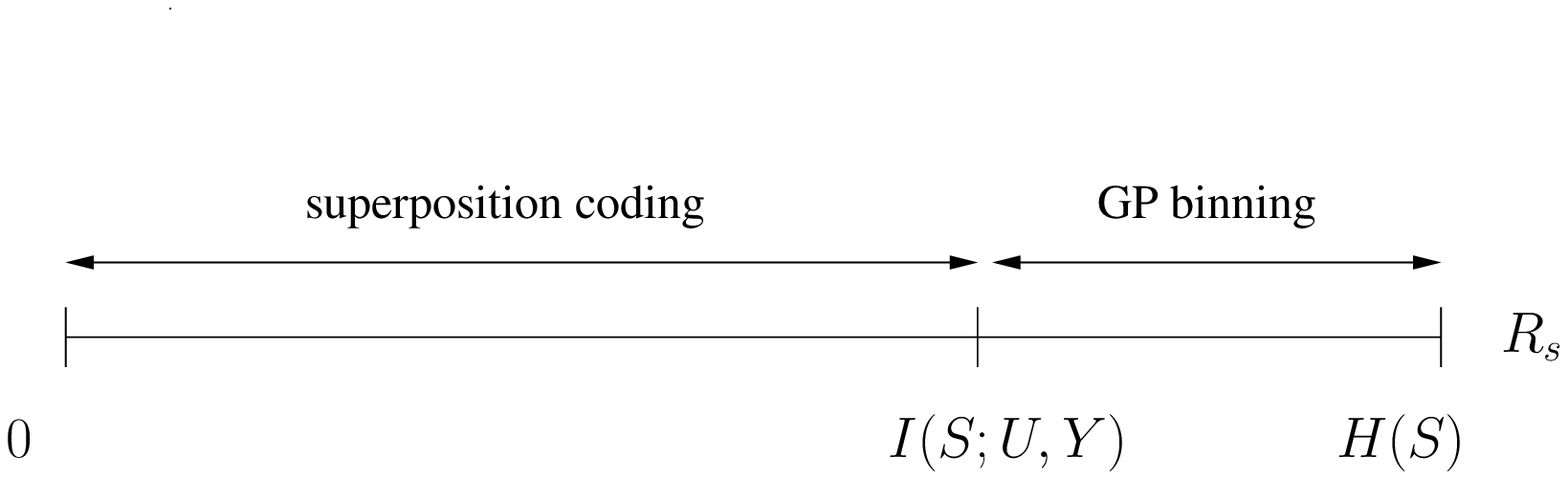,width=3.5in} \caption[Interference
channel.]{Binning against a codebook.}
      \label{f:BinningAgainstCodebook}
\end{figure}

%%%%%%%%%%%%%%%%%%
For the communication between the cognitive transmitter and its corresponding receiver, a codebook carrying $W_2$ creates
 interference. The situation is depicted  in
Fig~\ref{X2asinterference}, where $S$ plays the role of the codebook of rate $R_s$ interfering with the communication of message $W$ at rate $R$.
 %The encoding approach of
%Thm.~\thmref{jointdecoding} uses Gel'fand-Pinsker (G-P) approach to precode
%against $(X_{2a}, X_{2b})$.
While in the GP problem
the interference $S$ is generated by a discrete memoryless source (DMS), the interference in the
cognitive setting is a {\it codebook} of some rate, $R_s$. The next lemma reflects the fact that when $R_s$ is small, this can  be exploited for potential rate gains.
\begin{MyLemma} \lemmalabel{GerhardLemma}
For the communication situation of Fig.~\ref{f:BinningAgainstCodebook}, the rate
\begin{align}
%R \le &\max_{P_{U,X|S}} \min \{ I(X;Y|S), \nonumber \\
R \le &\max_{P_{U|S}, f(\cdot)} \min \{ I(X;Y|S), \nonumber \\
& \max \{ I(U,S;Y)-R_s, I(U;Y)-I(U;S) \} \} \eqnlabel{rateimprovement}
\end{align}
is achievable.

For $I(S;U,Y) \le R_s \le H(S)$, binning achieves the GP rate given by the second term in \eqnref{rateimprovement}.

For  $R_s \le I(S;U,Y)$, superposition coding achieves the rate given by the first term in \eqnref{rateimprovement}.
\end{MyLemma}
The two cases are shown in Fig.~\ref{f:BinningAgainstCodebook}.
%%%%%%%%%%%%%%%%%%%%%%%%%%%%%%%%%%%%%%%%%%%%
%%%%%%%%%%%%%%%%%%%%

%%%%%%%%%%%%%%%%%%%%%%%%%%%%%%%%%%%%%%%%%%%%%%%%%%%%%%%%%%%
\begin{proof}
See Appendix B.
\end{proof}
%%%%%%%%%%%%%%%%%%%%%%%%%%%%%%%%%%%%%%%%%%%%%%%%%%%%%%%%%%%
%%%%%%%%%%%%%%%%%%%%%%%%%%%%%%%%
 \begin{MyRemark}
 Rate \eqnref{rateimprovement} can be written as
 \begin{align}
R \le &  \max_{P_{U|S}, f(\cdot)} \{ I(X,S; Y) \nonumber \\
 &- \max \{ I(S;Y) \min \{ R_s, I(U, Y; S) \} \}. \eqnlabel{rateimprovement2}
\end{align}
 \end{MyRemark}
 %%%%%%%%%%%%%%%%%%%%%%%%%%%%%%
 \vspace{0.1in}
 From \eqnref{rateimprovement} and \eqnref{rateimprovement2}, we observe that $I(S;U,Y) \le R_s \le H(S)$, corresponds to the classical GP setting. Potential rate improvement comes for $R_s \le I(S;U,Y)$. Interestingly, in this case the receiver decodes both indexes $(w,j)$, thus learning both its message and the interference.
 A related setting in which both data and the channel state information is communicated to the receiver was analyzed in \cite{CoverKimSutivong2007, CoverKimSutivongArXiv}.

 In the cognitive setting of Fig.~\ref{f:Figure}, index $j$ carries information about $W_2$. The implication is that, when $R_s$ is small, receiver $1$ will decode a part (or the whole) of $W_2$ without having encoder $2$ rate split to send common information in the sense of \cite{Carleial78, HanKobayashi81}.

%Due to Lemma~\lemmaref{GerhardLemma}, rates of
%Thm.~\thmref{jointdecoding}  can be improved.
Recall that, due to rate-splitting, encoder $1$ uses two codebooks to send a common and a private index. We denote these respective codebooks as $(U_{1c}^N, U_{1a}^N)$.
We can distinguish four cases depending on whether the two codebooks are generated through binning or superposition coding:
\begin{enumerate}
\item Binning: Both $(U^N_{1c}, U^N_{1a})$ are binned against the codebook of the non-cognitive encoder, $X^N_{2}$.
\item Superposition coding: Codebooks are superimposed on $X_2^N$.
\item Binning then superposition coding: $U_{1c}^N$ is binned against $X_2^N$, and $U_{1a}^N$ is superimposed on $(X^N_2, U^N_{1c})$.
\item Superposition coding then binning: $U_{1c}^N$ is superimposed on $X_2^N$; $U_{1a}^N$ is superimposed on $U^N_{1c}$ and binned against $X^N_2$.
\end{enumerate}
%The relationships between the random variables are shown in Fig.?
In the last two cases, decoder $1$ can decode $W_2$ due to superposition coding of $U^N_{1a}$ or $U^N_{1c}$ on $X^N_2$,  as shown in Lemma~\lemmaref{GerhardLemma}. The setting thus corresponds to the {\it cognitive radio with the degraded message set}. For this channel model, superposition coding achieves the capacity \cite{JiangXinGarg2007, LiangBaruchPoorShamaiVerdu2007}. The two last cases can therefore bring no improvement.
The achievable rate region is the union of two rate regions, achieved by binning or superposition coding. We will derive these regions after formally defining the problem in the next section.
We remark that in the above encoding scheme, codebook $U_{1a}^N$  is always superimposed on  $U_{1c}^N$. The other encoding choice would be to use binning for $U_{1a}^N$ against the codebook carrying the common message, $U_{1c}^N$.

  As the final point about the proposed scheme we note that encoder $2$ also uses rate-splitting and forms two codebooks $(X^N_{2a}, X^N_{2b})$ using superposition coding. Encoder $1$ is binning against both codebooks and is not decoding a part of $W_2$. An interesting next step would therefore be to choose respective rates $R_{2a}$ and $R_{2b}$ following Lemma~\lemmaref{GerhardLemma}  such that $(U^N_{1a}, U^N_{1c})$ are binned against one of the two codebooks, but superimposed on the other. That would facilitate decoding a part of $W_2$ at receiver $1$.

%%%%%%%%%%%%%%%%%%%%%%%%%%%%%%%%%%%%%%%%%%%%%%%%%%%%%%%%%%%%%%%%%%%%%%%%%%%%%%
%For the first case, the rates are given by Thm.~\thmref{jointdecoding}.
%The achievable rates are stated in Thms?-?
%\begin{MyTheorem}\thmlabel{jointdecodingbinning}
%%%%%%%%%%%%%%%%%%%%%%%%%%%
%\begin{align}
%&R_{1a} \le \min \{I(U_{1a};Y_1|X_{2a}, X_{2b}, U_{1c},Q), \nonumber \\
% & \max \{I(U_{1a};Y_1|U_{1c},Q)- I(U_{1a}; X_{2a}, X_{2b}|U_{1c}, Q), \nonumber \\
% & I(U_{1a}, X_{2a},X_{2b};Y_1|U_{1c},Q)- R_2 \} \}  \eqnlabel{R1abinningN}  \\
%& R_1    \le \min \{ I(U_{1a}, U_{1c};Y_1|X_{2a}, X_{2b}, Q)
%\nonumber \\
%& \max \{I(U_{1c},U_{1a}; Y_1|Q)   - I(U_{1c}, U_{1a};
%X_{2a}, X_{2b}|Q) \nonumber \\
%& I(U_{1a},U_{1c}, X_{2a},X_{2b};Y_1|,Q)- R_2 \} \} \\
%&R_2 \le I(X_{2a}, X_{2b}; Y_2, U_{1c}|Q) \eqnlabel{jdR2aR2b1N}\\
%&R_2 + R_c \le I(X_{2a},X_{2b},U_{1c}; Y_2|Q) \\
%&R_{2b} \le I(X_{2b}; Y_2, U_{1c}|X_{2a},Q) \eqnlabel{jdR2N}\\
%&R_{2b} + R_c \le I(X_{2b},U_{1c}; Y_2|X_{2a},Q) \eqnlabel{jdR2bRcN}
%\end{align}
%for some joint distribution that factors as
%$p(q)p(x_{2a},x_{2b},u_{1c},u_{1a},x_1,x_2|q) p(y_1,y_2|x_1,x_2)$
%and for which all the right-hand sides are nonnegative.
%\end{MyTheorem}
%%%%%%%%%%%%%%%%%%%%%%%%%%%%%%%%%%%%%%%%
\section{Channel Model}
%%%%%%%%%%%%%%%%%%%%%%%%%%%%%%%%%%%%%%%%
Consider a channel with finite input alphabets $\Xc_1, \Xc_2$,
finite output alphabets $\Yc_1,\Yc_2$, and a conditional probability
distribution $p(y_1,y_2|x_1, x_2)$, where $(x_1,x_2) \in {\Xc}_1
\times {\Xc}_2$ are channel inputs and $(y_1,y_2) \in {\Yc}_1 \times
{\Yc}_2$ are channel outputs.  Each encoder $t$, $t=1,2$, wishes to
send a message $W_t \in \{1, \ldots, M_t\}$ to decoder $t$ in $N$
channel uses. Message $W_2$ is also known at encoder $1$ (see
Fig.~\ref{f:Figure}). The channel is memoryless and time-invariant
in the sense that
\begin{align}
p(&y_{1,n},y_{2,n}|x^n_1,x^n_2,y^{n-1}_1,y^{n-1}_2, {\bar
w}) \nonumber\\
&=p_{Y_1,Y_2|X_1,X_2}(y_{1,n},y_{2,n}|x_{1,n},x_{2,n})
\eqnlabel{channel}
\end{align}
for all $n$, where $X_1,X_2$ and $Y_1,Y_2$ are random variables
representing the respective inputs and outputs, ${\bar w} =
[w_1,w_2]$ denotes the messages to be sent, and
$x_t^n=\begin{bmatrix} x_{t,1},& \ldots, &x_{t,n} \end{bmatrix}$. We
will follow the convention of dropping subscripts of probability
distributions if the arguments of the distributions are lower case
versions of the corresponding random variables.

%%%%%%%%%%%%%%%%%%
% OUR PAPER
%%%%%%%%%%%%%%%%%%
%%%%%%%%%%%%%%%%%%%%%%
% CODE FOR THE CHANNEL
%%%%%%%%%%%%%%%%%%%%%%
An $(M_1,M_2, N,P_e)$ code has two encoding functions
\begin{align}
X^N_1& =f_1(W_1,W_2) \eqnlabel{x1} \\
X^N_2& =f_2(W_2) \eqnlabel{x2}
\end{align}
 two decoding functions
\begin{equation} \eqnlabel{decodingfunctions}
\hat W_t=g_t(Y^N_t) \quad t=1,2
\end{equation}
%%%%%%%%%%%%%%%%%%%%%%%%%%%%
% PROB. OF ERROR OF THE CODE
%%%%%%%%%%%%%%%%%%%%%%%%%%%%
and an error probability
\begin{equation}\eqnlabel{Pe}
P_e= \max\{ P_{e,1}, P_{e,2}   \}
\end{equation}
where, for $t=1,2$, we have
\begin{equation} \eqnlabel{Pet}
P_{e,t}= \sum_{(w_1,w_2)}\frac{1}{M_1M_2} P\left[ g_t(\Yv_t) \neq
 w_t| (w_1,w_2) \mbox{ sent}  \right ].
\end{equation}
%%%%%%%%%%%%%%%%%%%%%
% ACHIEVABLE RATES
%%%%%%%%%%%%%%%%%%%%%
A rate pair $(R_1,R_2)$ is achievable if, for any $\epsilon > 0$,
there is an $(M_1, M_2, N,P_e)$ code such that
\begin{displaymath}
 M_t \ge 2^{NR_t}, \quad t=1,2, \mbox{ and } P_e \le \epsilon.
\end{displaymath}
%%%%%%%%%%%%%%%%%%
%%%%%%%%%%%%%%%%%%%%%%%%%%%%%%%
% CAPACITY REGION
%%%%%%%%%%%%%%%%%%%%%%%%%%%%%%%%
The capacity region of the interference channel with a cooperating
encoder is the closure of the set of all achievable rate pairs
$(R_1,R_2)$.
%%%%%%%%%%%%%%%%%%%%%%%%%%%%%%%%%%%%%%%%%%%%%%%%%%%%%%%%%%%%%%%%%%%%%%%%%%%%%%%%%%%%%%%%%%%%%%%%%%%
\section{Achievable Rate Region} \label{sect:achievablerates}
%%%%%%%%%%%%%%%%%%%%%%%%%%%%%%%%%%%%%%%%%%%%%%%%%%%%%%%%%%%%%%%%%%%%%%%%%%%%%%%%%%%%%%%%%%%%%%%%%%%
%%%%%%%%%%%%%%%%%%%%%%%%%%%%%%%%%%%%%%%%%%%%%%%%%%%%%%%%%%%%%%%%%%%%%%%%%%%%%%%%%%%%%%%%%%%%%%%%%%%
\begin{figure}[t]
\center\epsfig{figure=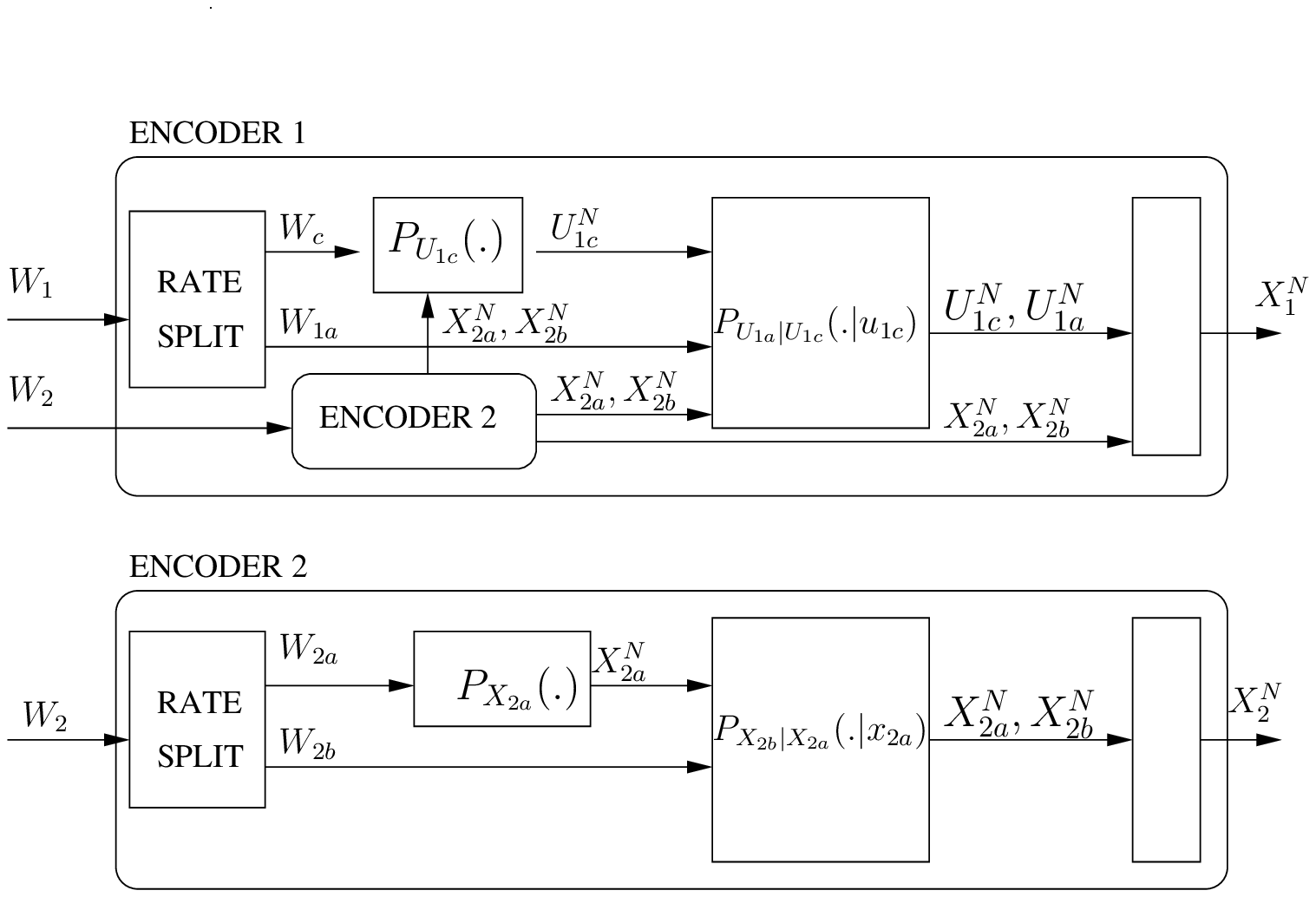,width=3.5in}
\caption[Interference channel with unidirectional
cooperation.]{Encoding structure.}
      \label{f:encodingstructure}
\end{figure}
%%%%%%%%%%%%%%%%%%%%%%%%%%%%%%%%%%%%%%%%%%%%%%%%%%%%%%%%%%%%%%%%%%%%%%%%%%%%%%%%%%%%%%%%%%%%%%%%%%%%%
To obtain an inner bound, we employ rate splitting. We let
\begin{align}
R_1& = R_{1a}+R_c \eqnlabel{ratesplitting1} \\
R_2& = R_{2a}+R_{2b} \eqnlabel{ratesplitting2}
\end{align}
for nonnegative $R_{1a}, R_c, R_{2a}, R_{2b}$ which we now specify.

%%%%%%%%%%%%%%%%%%%%%%%%%
% INSIGHT INTO THE SCHEME
%%%%%%%%%%%%%%%%%%%%%%%%%
In the encoding scheme, encoder $2$  uses superposition coding with
two codebooks $X^N_{2a}, X^N_{2b}$. Encoder $1$ repeats the steps of
encoder $2$ and adds binning: it encodes the split message $W_1$
with two codebooks which are Gel'fand-Pinsker precoded against
$X^N_{2a}, X^N_{2b}$. In particular:
\begin{enumerate}
\item Binning against $X^N_{2a}, X^N_{2b}$ is used to create a codebook $U^N_{1c}$ of common
rate $R_c$.
\item Binning against $X^N_{2a}, X^N_{2b}$ conditioned on $U_{1c}$ is used to create a codebook $U^N_{1a}$
with private rate $R_{1a}$.
\end{enumerate}
%%%%%%%%%%%%%%%%%%%%%%%%%%%%%%%%%%%%%%%%%%%%%%%%%%%%%%%%%%%%%%%%
The encoding structure is shown in Fig.~\ref{f:encodingstructure}.

We have the following result.
%%%%%%%%%%%%%%%%%%%%%%%%%%%%%%%%%%%%%%%%%%%%%%%%%%%%%
\begin{MyTheorem}\thmlabel{jointdecoding} {\em(joint decoding)}
Rates \eqnref{ratesplitting1}-\eqnref{ratesplitting2} are achievable
if
%%%%%%%%%%%%%%%%%%%%%%%%%%
\begin{align}
&R_{1a} \le I(U_{1a}; Y_1|U_{1c}, Q)   -I(U_{1a};X_{2a},X_{2b}|U_{1c},Q) \eqnlabel{jdR1a}  \\
& R_1   \le  I(U_{1c},U_{1a}; Y_1|Q)
   - I(U_{1c}, U_{1a};
X_{2a}, X_{2b}|Q) \eqnlabel{jdR1aRc}\\
&R_2 \le I(X_2; Y_2, U_{1c}|Q) \eqnlabel{jdR2aR2b1}\\
&R_2 + R_c \le I(X_2,U_{1c}; Y_2|Q) \\
&R_{2b} \le I(X_{2b}; Y_2, U_{1c}|X_{2a},Q) \eqnlabel{jdR2}\\
&R_{2b} + R_c \le I(X_{2b},U_{1c}; Y_2|X_{2a},Q) \eqnlabel{jdR2bRc}
\end{align}
for some joint distribution that factors as
\begin{equation}
p(q)p(x_{2a},x_{2b},u_{1c},u_{1a},x_1,x_2|q) p(y_1,y_2|x_1,x_2) \eqnlabel{jointdistr}
\end{equation}
and for which all the right-hand sides are nonnegative.
\end{MyTheorem}
\begin{proof}
See Appendix A.
\end{proof}
 %%%%%%%%%%%%%%%%%%%%%%%%%%%%%%%%%%%%%%%%%%%%%%%%%%%%
 \begin{MyTheorem}\thmlabel{sequentialdecoding} {\em (sequential decoding)}
Rates \eqnref{ratesplitting1}-\eqnref{ratesplitting2} are achievable
if
%%%%%%%%%%%%%%%%%%%%%%%%%%%%%%%%%%%%%%%%%%%%%%%%
\begin{align}
&R_{1a} \le I(U_{1a}; Y_1|U_{1c}, Q) -I(U_{1a};X_2|U_{1c},Q) \eqnlabel{achR1a}  \\
&R_c  \le \min \{ I(U_{1c}; Y_1|Q), I(U_{1c};Y_2, X_{2a}|Q) \} \nonumber\\
& \qquad - I(U_{1c};X_2|Q)
\eqnlabel{achRc}\\
&R_{2a} \le I(X_{2a};Y_2|Q) \eqnlabel{achR2a}\\
&R_{2b} \le I(X_{2b};Y_2,U_{1c}|X_{2a},Q) \eqnlabel{achR2b}
\end{align}
for some joint distribution that factors as
$p(q)p(x_{2a},x_{2b},u_{1c},u_{1a},x_1,x_2|q) p(y_1,y_2|x_1,x_2)$
and for which the right-hand side of \eqnref{achR1a} and
\eqnref{achRc} are nonnegative.
 $Q$ is a time sharing random
variable.
\end{MyTheorem}
\begin{proof}
The proof follows similar steps as the Thm.~\thmref{jointdecoding} proof and is omitted. Details can be found in \cite{MGKSITA2007}.
\end{proof}
\begin{MyRemark} The rates of Thm.~\thmref{jointdecoding} include
the rates of Thm.~\thmref{sequentialdecoding}.
\end{MyRemark}
%%%%%%%%%%%%%%%%%%%%%%%%%%%%%%%%%%%%%%%%%%%%%%%%%%%%%
% COMPARISON
%%%%%%%%%%%%%%%%%%%%%%%%%%%%%%%%%%%%%%%%%%%%%%%%%%%%%
\begin{MyRemark}
Thm.~\thmref{jointdecoding}  includes the rates of the following
schemes:
\begin{itemize}
\item The scheme of  \cite[Thm
$3.1$]{WuVishwanath06} for $X_{2a} = \emptyset, U_{1c} = \emptyset,
X_{2b} = (X_2, U)$ and $U_{1a}=V$ achieving:
\begin{align}
R_2 &\le I(X_2,U;Y_2) \\
R_1 & \le I(V;Y_1) - I(V;X_2,U)
\end{align}
for $p(u,x_2)p(v|u,x_2)p(x_1|v)$.
\item The scheme of \cite[Lemma $4.2$]{JovicicViswanath06Submitted} for
$X_{2a} = \emptyset$, $X_{2b} = X_2$, $U_{1a}= \emptyset$, and
$R_1=R_c$, $R_2=R_{2b}$ as:
\begin{align}
R_2 &\le I(X_2;Y_2| U_{1c})\nonumber \\
R_1 & \le \min \{ I(U_{1c}; Y_1), I(U_{1c};Y_2) \} \nonumber
\end{align}
for $p(x_2)p(u_{1c})$.   The strategy in
\cite{JovicicViswanath06Submitted} considers  the case when
$I(U_{1c};Y_1) \le I(U_{1c};Y_2)$.
\item Carbon-copy on dirty paper \cite{KhistiErez06}  for  $X_{2a} = \emptyset, U_{1a} =
\emptyset$.
%\item The strong interference case \cite{MYKsubmitted,MYKSD2006} for $X_{2a}=X_2, X_{2b} = \emptyset, U_{1c}=X_2, U_{1a}=X_1.$
\item For $X_{2a}=\emptyset$, our scheme closely resembles the scheme in
\cite{JiangXin2007}. The first difference in our scheme is that two
binning steps are not done independently which brings potential
improvements. The second difference is in the evaluation of error
events.
\end{itemize}
%%%%%%%%%%%%%%%%%%%%%%%%%%%%%%%%%%%%%%%%%%%%%%%%%%%%%%%
%%%%%%%%%%%%%%%%%%%%%%%%%%%%%%%%%%%%%%%%%%%
% 1. COMPARISON TO DEVROYE AND XIN
%%%%%%%%%%%%%%%%%%%%%%%%%%%%%%%%%%%%%%%%%%%
 It is also interesting to compare our scheme to the
encoding scheme in \cite{DevroyeSubmitted}. The latter combines rate
splitting at both users,  with two-step binning at the cognitive
user.  Each user sends a private index decoded by its receiver, and
a common index decoded by both. Again, one difference in our scheme
is that two binning steps are not independent. The other is that in
our scheme the cognitive encoder cooperates by encoding index $W_2$.
%And while it is not
%obvious that, in our scheme, a common message is sent from encoder
%$2$ to both receivers, we note that decoder $1$ actually decodes
%index $W_{2a}$ (or $W_{2b}$) by decoding a bin because each bin
%corresponds to a different codeword pair.
%PROVE!
\end{MyRemark}
The next rate region is obtained by exploiting Lemma~\lemmaref{GerhardLemma}.
%%%%%%%%%%%%%%%%%%%%%%%%%%%%%%%%%%%%%%%%%%%%%%%%%%%%%%%%%%%%%%%%%
\subsection{An Achievable Rate Region with Superposition Coding}
%%%%%%%%%%%%%%%%%%%%%%%%%%%%%%%%%%%%%%%%%%%%%%%%%%%%%%%%%%%%%%%%%
Consider a joint distribution \eqnref{jointdistr} and rate $R_2$ that satisfy
\begin{align}
R_2 &\le I(X_2; U_{1c}, Y_1) \eqnlabel{cond1} \\
R_2 &\le I(X_2; U_{1a}, Y_1|U_{1c}). \eqnlabel{cond2}
\end{align}
   From Lemma~\lemmaref{GerhardLemma}, we know that under respective conditions \eqnref{cond1} and \eqnref{cond2}, superposition of $U_{1c}^N$ and $U_{1a}^N$ with  $X_2^N$ should be used instead of binning. The encoding scheme of the cognitive encoder reduces to rate-splitting and superposition coding. The scheme and the obtained rates reduce to that of \cite[Thm.$5$]{JiangXin2007} derived for the {\it cognitive radio with the degraded message set}, in which the cognitive decoder needs to decode both messages. No rate-splitting at encoder $2$ is needed. We restate the result for completeness.

%%%%%%%%%%%%%%%%%%%%%%%%%%%%%%%%%%%%%%%%%%%
   Achievable rates $(R_1,R_2)$ satisfy
 \begin{align}
 R_{1a} & \le I(X_1; Y_1| X_2, U_{1c}) \nonumber \\
 R_1 & \le I(X_1;Y_1|X_2)  \nonumber \\
 R_1 +R_2 & \le I(X_1, X_2; Y_1) \nonumber \\
 R_c+R_2 & \le I(U_{1c},X_2;Y_2) \eqnlabel{superposition}
 \end{align}
 for some joint input distribution $p(x_2,u_{1c},x_1)$.
%%%%%%%%%%%%%%%%%%%%%%%%%%%%%%%%%%%%%%%%%%%%%%

After Fourier-Motzkin elimination \cite{Lall2004}, the rates \eqnref{superposition} reduce to the following region.
   \begin{MyTheorem} \cite{JiangXinGarg2007}.
   Achievable rates $(R_1,R_2)$ satisfy
 \begin{align}
 R_1 & \le I(X_1;Y_1|X_2) \nonumber \\
 R_2 & \le I(U_{1c},X_2;Y_2) \nonumber \\
 R_1 +R_2 & \le I(X_1; Y_1|X_2,U_{1c}) + I(U_{1c},X_2; Y_2) \nonumber \\
 R_1+R_2 & \le I(X_1,X_2;Y_1)
 \end{align}
 for some joint input distribution $p(x_2,u_{1c},x_1)$.
\end{MyTheorem}
\begin{MyRemark}
The above region is the capacity region for the cognitive radio with  degraded message sets: the converse follows from \cite{LiangBaruchPoorShamaiVerdu2007} where a more general case of confidential messages is analyzed. The result follows by considering the special case of no security.
\end{MyRemark}

We have so far presented achievable rates for the cognitive radio channel.
We next derive two outer bounds to performance of any encoding scheme in this channel.
%%%%%%%%%%%%%%%%%%%%%%%%%%%%%%%%%%%%%%%%
\section{Outer Bounds} \label{sect:outerbounds}
%%%%%%%%%%%%%%%%%%%%%%%%%%%%%%%%%%%%%%%%
\begin{MyTheorem} \thmlabel{generalouterbound}
The set of rate pairs $(R_1,R_2)$ satisfying
\begin{align}
R_1& \le I(V,U_1;Y_1) \eqnlabel{R1u2}\\
R_2& \le I(V,U_2;Y_2) \eqnlabel{R2u2}  \\
R_1+R_2& \le \min \{ I(V,U_1;Y_1)+I(U_2;Y_2|U_1,V), \eqnlabel{R1R20u2} \\
& \qquad \quad I(U_1;Y_1|U_2,V)+ I(V,U_2;Y_2) \} \eqnlabel{R1R2u2}
\end{align}
for input distributions $p(v,u_1,u_2,x_1,x_2)$ that factor as
\begin{equation} \eqnlabel{pdf}
%p(u_1)p(u_2)p(v|u_1,u_2)p(x_2|v,u_2)p(x_1|v,u_1,u_2,x_2)
p(u_1)p(u_2)p(v|u_1,u_2)p(x_2|u_2)p(x_1|u_1,u_2)
\end{equation}
is an outer bound to the capacity region.
\end{MyTheorem}
%%%%%%%%%%%%%%%%%%%%%%%%%%%%%%%%%%%%%%%%%%%
%%%%%%%%%%%%%%%%%%
\begin{proof}
See Appendix C.
\end{proof}
%%%%%%%%%%%%%%%%%%
\begin{MyRemark} We observe that \eqnref{R1u2}-\eqnref{R1R2u2} is of
the same form as the outer bound for the broadcast channel in
\cite[Sect. $3$]{NairElGamal2007}. The
 difference is the factorization of the input distribution.
\end{MyRemark}
\begin{MyRemark} One can restrict attention to distributions
\eqnref{pdf}
 where $X_2$ is a function of $U_2$ and $X_1$ is a function of
 $(U_1,U_2)$.
 The bounds \eqnref{R1u2}-\eqnref{R1R2u2}
 can then be written as
 \begin{align}
R_1& \le I(V,U_1;Y_1) \eqnlabel{R1u2new}\\
R_2& \le I(V,U_2,X_2;Y_2) \eqnlabel{R2u2new}  \\
R_1+R_2& \le \min \{ I(V,U_1;Y_1)+I(X_1,X_2;Y_2|U_1,V), \eqnlabel{R1R20u2new} \\
& \qquad \quad I(X_1;Y_1|X_2,U_2,V)+ I(V,U_2,X_2;Y_2) \}
\eqnlabel{R1R2u2new}
\end{align}
From \eqnref{R2u2new} and \eqnref{R1R2u2new}, we obtain the outer bound of
\cite[Thm. $3.2$]{WuVishwanath06}:
 \begin{align}
R_2& \le I(X_1;Y_1|X_2) \eqnlabel{R2weak}\\
R_2& \le I(U,X_2;Y_2) \eqnlabel{R2u2weak}  \\
R_1+R_2& \le   I(X_1;Y_1|X_2,U)+ I(U,X_2;Y_2) \eqnlabel{R1R2u2weak}
\end{align}
where we used notation $U=[U_2,V]$ and also added \eqnref{R2weak}\
as it follows by standard methods. The probability distribution
factors as
\begin{equation}
p(u,x_1,x_2) p(y_1,y_2|x_1,x_2). \eqnlabel{pdfweak}
\end{equation}
  Interestingly, \eqnref{R2weak}-\eqnref{pdfweak} was shown to be tight under weak interference
   \cite[Def. $2.3$]{WuVishwanath06} and in
particular
   for Gaussian channels with weak
  interference \cite{WuVishwanath06, JovicicViswanath2006}.
\end{MyRemark}

%%%%%%%%%%%%%%%%%%%%%%%%%%%%%%%%%%%%%%%%%%%%%%%%%%%%%%%%%%%%%%%%%
 The following theorem gives a simple upper bound in strong interference.
 %%%%%%%%%%%%%%%%%%%%%%%%%%%%%%%%%%%%%%%%%%%%%%%%%%%%%%%%%%%%%
\begin{MyTheorem} \thmlabel{stronginterferenceouterbound}
For an interference channel with one cooperating encoder satisfying
\begin{equation}\eqnlabel{decodeboth}
I(X_1;Y_1|X_2) \le I(X_1;Y_2|X_2)
\end{equation}
 for all input distribution $p(x_1,x_2)$, the set of
rate pairs $(R_1,R_2)$ satisfying
\begin{align}
R_1& \le I(X_1;Y_1|X_2) \eqnlabel{R1si}\\
R_1+R_2& \le  I(X_1,X_2;Y_2)\eqnlabel{R1R2si}
\end{align}
for all input distributions $p(x_1,x_2)$ is an outer bound to the
capacity region.
\end{MyTheorem}
\begin{proof}
See  Appendix D.
\end{proof}
\begin{MyRemark} The bound \eqnref{R1R2si} reflects the fact that, because decoder $2$
experiences strong interference, as given by \eqnref{decodeboth}, it can decode $W_1$ with no rate
penalty.
\end{MyRemark}
We next compare the outer bound of Thm.~\thmref{stronginterferenceouterbound} to the achievable rates for Gaussian channels.
%%%%%%%%%%%%%%%%%%%%%%%%%%%%%%%%%%%%%%%%
\section{Gaussian Channel} \label{sect:Gaussianchannel}
%%%%%%%%%%%%%%%%%%%%%%%%%%%%%%%%%%%%%%%%
To illustrate obtained results more concretely, we next consider the Gaussian interference channel described by
\begin{align}
Y_{1}=X_1+a X_2+Z_1   \eqnlabel{channel1} \\
Y_{2}=bX_1+ X_{2}+Z_{2} \eqnlabel{channel2}
\end{align}
where $Z_t \sim [0,1]$ and $E[X_t^2] \le P_t$, $t=1,2$. In the case
of weak interference, i.e., $b \le 1$, the capacity region was
determined in \cite{WuVishwanath06,JovicicViswanath2006}.
 %%%%%%%%%%%%%%%%%%%%%%%%%%%%%%%%%%%%%

We next evaluate the rates of Thm.~\thmref{jointdecoding} for the special case $X_{2a}=\emptyset$ and $Q=\emptyset$. Rates of
Thm.\thmref{jointdecoding} for this case  reduce to
%%%%%%%%%%%%%%%%%%%%%%%%%%
\begin{align}
&R_{1a} \le I(U_{1a}; Y_1|U_{1c})
  -I(U_{1a};X_2|U_{1c})  \nonumber \\
& R_1    \le  I(U_{1c},U_{1a}; Y_1)
   - I(U_{1c}, U_{1a}; X_2) \nonumber \\
&R_2  \le I(X_2; Y_2, U_{1c}) \nonumber \\
& R_2 + R_c \le I(X_2,U_{1c}; Y_2). \eqnlabel{QX2a0}
\end{align}
To simplify \eqnref{QX2a0}, we express the conditional entropies in terms of joint entropies, recall
that $R_1=R_c+R_{1a}$, and apply Fourier-Motzkin elimination to obtain
%%%%%%%%%%%%%%%%%%%%%%%%%%
\begin{align}
& R_1    \le  I(U_{1c},U_{1a}; Y_1)
   - I(U_{1c}, U_{1a}; X_2) \nonumber\\
&R_2  \le I(X_2; Y_2, U_{1c}) \nonumber\\
& R_2  \le I(X_2,U_{1c}; Y_2) \nonumber \\
& R_1 + R_2 \le I(X_2,U_{1c}; Y_2)+I(U_{1a}; Y_1, U_{1c}) \nonumber
\\
 & \qquad \quad -I(U_{1a};X_2, U_{1c}).
  \eqnlabel{G2}
\end{align}

%%%%%%%%%%%%%%%%%%%%%%%%%%%%%%%%%%%%%%%%%%%%%%%%%%%
%\subsection{When Sequential Decoding is Enough}
%%%%%%%%%%%%%%%%%%%%%%%%%%%%%%%%%%%%%%%%%%%%%%%%%%%
It is interesting to evaluate the rates of Thm.~\thmref{sequentialdecoding} achieved with sequential decoding for $X_{2a}=\emptyset, Q=\emptyset$ as was done for joint decoding in \eqnref{QX2a0}. This evaluation results in
\begin{align}
&R_{1a} \le I(U_{1a}; Y_1|U_{1c}) -I(U_{1a}; X_2|U_{1c}) \nonumber  \\
&R_c  \le \min \{ I(U_{1c}; Y_1), I(U_{1c};Y_2) \}  - I(U_{1c}; X_2)
\nonumber \\
&R_2 \le I(X_2;Y_2,U_{1c}). \eqnlabel{Gseq}
\end{align}
\begin{MyRemark}
When $I(U_{1c};Y_1) \le I(U_{1c};Y_2)$, decoder $2$ can decode  $W_1$. Thus, there is no need to rate split at encoder $1$ and we choose $U_{1a}=\emptyset$. It follows from \eqnref{G2} and \eqnref{Gseq} that for this case the same rates can be achieved by sequential decoding or by joint decoding.
\end{MyRemark}
\begin{MyRemark}
We observe from \eqnref{Gseq} that $R_c$, being a common rate, is bounded by the worst channel, as reflected by the  $\min\{I(U_{1c};Y_1),I(U_{1c};Y_2) \}$ term. If $I(U_{1c};Y_1) > I(U_{1c};Y_2)$, transmitting  $X_{2a}$ will allow decoder $2$ to decode part of $W_2$ before decoding $W_c$. It will also serve as an observation when decoding $W_c$ as suggested by the expression $I(U_{1c};Y_2,X_{2a})$ in \eqnref{achRc}. This will improve the common rate $R_c$.
\end{MyRemark}
%%%%%%%%%%%%%%%%%%%%%%%%%%%%%%%%%%%%%%%%%%%%%%%%%%%%

We evaluated region \eqnref{G2} for
\begin{displaymath}
X_2 \sim \Nc[0,P_2], \quad X_{1c} \sim \Nc[0,\alpha \beta P_1],
\quad X_{1a} \sim \Nc[0,\alpha \bar \beta P_1]
\end{displaymath}
\begin{align}
&U_{1c} = X_{1c} + \lambda_1 X_2 \nonumber \\
&U_{1a} = X_{1a} + \lambda_2 X_2 \nonumber \\
&X_1 = X_{1c} + X_{1a} + \sqrt{\frac{\bar \alpha P_1}{P_2}} X_2 \eqnlabel{rvs}
\end{align}
where $\Nc[0, \sigma^2]$ denotes the normal distribution with variance $\sigma^2$, and $0 \le \alpha, \beta \le 1$ and $0 \le \lambda_1, \lambda_2$.
Parameters $\alpha$ and $\beta$ determine the amount of power that the cognitive user dedicates respectively for cooperation ($1-\alpha$) and for sending the common message.

We compared the achievable region \eqnref{G2} to the
outer bound of Thm.~\thmref{stronginterferenceouterbound} which in
Gaussian channels is given by the following corollary:
\begin{MyCorollary} \corlabel{Gaussouterbound}
When $b \ge 1$, any achievable rate pair $(R_1,R_2)$ satisfies
\begin{align}
R_1& \le C((1-\rho^2)P_1) \nonumber \\
R_1+R_2& \le  C(b^2P_1 + P_2+2 \rho
\sqrt{b^2P_1P_2})\eqnlabel{outerG}
\end{align}
for some $\rho$, $0 \le \rho \le 1$, where
\begin{equation}
C(x)=\frac{1}{2} \log(1+x).
\end{equation}
\end{MyCorollary}
\vspace{0.1in}

%%%%%%%%%%%%%%%%%%
% PLOTS
%%%%%%%%%%%%%%%%%%%%%%%%%%%%%%%%%%%%%%%%%%%%%%%%%%%%%%%%%%%%%%%%%%%%%%%%%%%%%%%%%%%%%%%%%%%%%%%%%%%
\begin{figure}[t]
\center\epsfig{figure=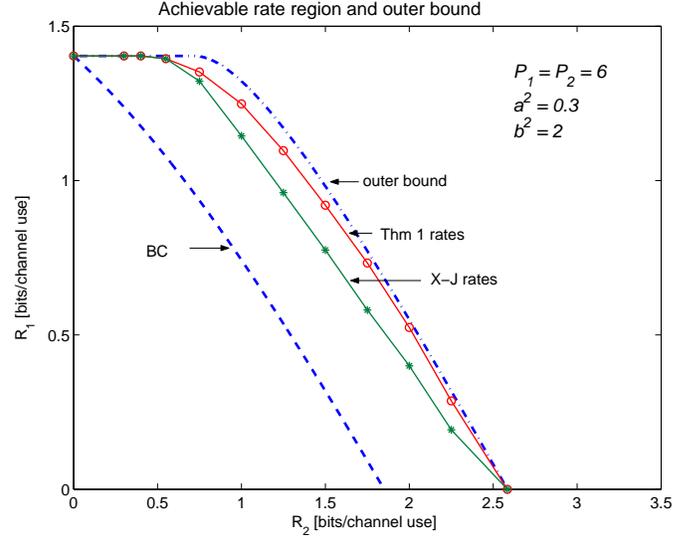,width=3.5in} \caption[Interference
channel with unidirectional cooperation.]{Achievable rates of
Thm.~\thmref{jointdecoding} and \cite[Thm.$5$]{JiangXin2007} and
outer bound of Cor.~\corref{Gaussouterbound}. Also shown is the
capacity region of a BC from the cooperative encoder, i.e. case $P_2=0$.
}
      \label{Plot3w}
\end{figure}
\begin{figure}[t]
%\center\epsfig{figure=Plot4.eps,width=3.5in}
\center\epsfig{figure=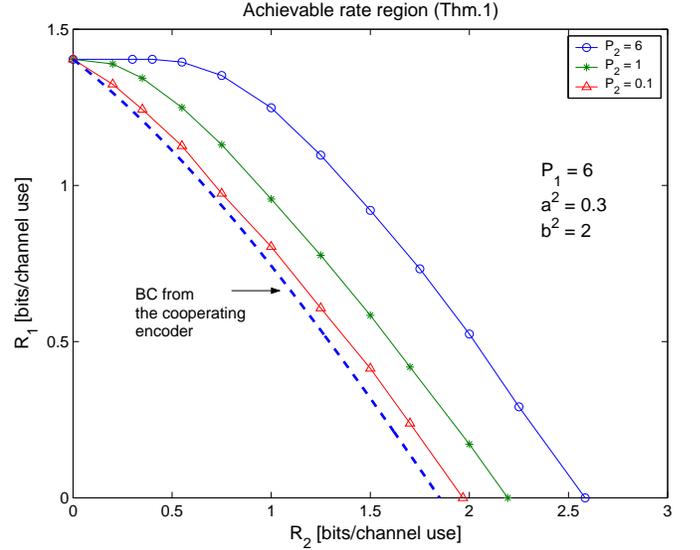,width=3.5in}
\caption[Interference
channel with unidirectional cooperation.]{Achievable rates for
different values of the non-cooperating encoder power, $P_2$.}
      \label{Plot4}
\end{figure}
\begin{figure}[t]
%\center\epsfig{figure=Plot6.eps,width=3.5in}
\center\epsfig{figure=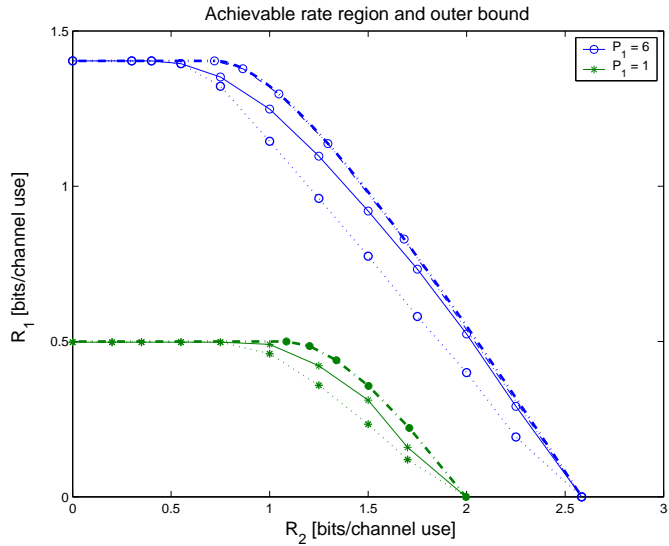,width=3.5in}
\caption[Interference
channel with unidirectional cooperation.]{Impact of reduced power of the
cognitive transmitter to achievable rates. Rates achieved with Thm.~$1$ are shown in solid lines and rates of \cite{JiangXin2007} are shown with dotted line. Dash-dotted line shows the outer bound.}
      \label{Plot6}
\end{figure}
%%%%%%%%%%%%%%%%%%%%%%%%%%%%%%%%%%%%%%%%%%%%%%%%%%%%%%%%%%%%%%%%%%%%%%%%%%%%%%%%%%%%%%%%%%%%%%%%%%%%%
Fig.~\ref{Plot3w} shows the achievable rate region \eqnref{G2} and the outer
bound \eqnref{outerG} for channel gain values
$a^2=0.3$, $b^2 = 2$ and equal powers $P_1=P_2=6$. We observe higher
rates of Thm.~\thmref{jointdecoding} compared to that of
\cite[Thm.~$5$]{JiangXin2007}.
%%%%%%%%%%%%%%%%%%%%%%%%%%%%%%%%%%%%%%%%%%%

When the encoder $2$ does not transmit (i.e. $P_2=0$), the channel reduces to the broadcast channel in which there is only the cooperating encoder communicating to the two receivers. The rates achieved in
the BC are also shown. Unlike the BC channel rate region, the
interference channel with one cooperating encoder region is flat for
smaller values of $R_2$,  reflecting the fact that for smaller values
of $R_2$ a cognitive transmitter does not need to cooperate. It can
instead use its full power to precode and transmit $W_1$ at the
single-user rate as if the second user was not present. It starts
cooperating only for higher $R_2$. At $R_1=0$, the cooperating encoder
fully helps encoder $2$, i.e. $\alpha=0$ and user $2$ benefits from the coherent combining gain as indicated by the rate expression
\begin{displaymath}
R_{2,max} = \frac{1}{2} \log \left(1 + \left(
1+b\sqrt{\frac{P_1}{P_2}} \right)^2P_2 \right).
\end{displaymath}
The achievable rates come very close to the outer bound, especially
for larger values of $R_2$, in the regime where the cognitive encoder dedicates more of its power to cooperate.

Fig.~\ref{Plot4} shows achievable rates for different values of power $P_2$ and fixed power $P_1$.  As $P_2$ decreases, the
performance gets closer to the rate achieved in the BC with only the cooperating encoder transmitting to the two receivers. Since in the BC  encoder $2$ is not present, the rate region does not depend on $P_2$ and is given by the dashed line.
Fig.~\ref{Plot6} shows the effect of reducing power at the
cognitive encoder, keeping $P_2$ constant. This has a higher impact, drastically reducing rate $R_1$.

%%%%%%%%%%%%%%%%%%%%%%%%%%%%%%%%%%%%%%%%%%
% Sequential Decoding
%%%%%%%%%%%%%%%%%%%%%%%%%%%%%%%%%%%%%%%%%%
For the Gaussian channel, the rates achieved with sequential encoding \eqnref{Gseq} can be evaluated for the choice of random variables $X_2, X_1, U_{1a}, X_{1a}$ as in \eqnref{rvs}. $U_{1c}$ carries a common message and is to be precoded against interference. Since the two channels from encoder $1$ to the two receivers experience different interference, the carbon-copy method of \cite{KhistiErez06} can be used.  More details on this approach are presented in \cite{MGKSITA2007}.

\begin{figure}[t]
\center\epsfig{figure=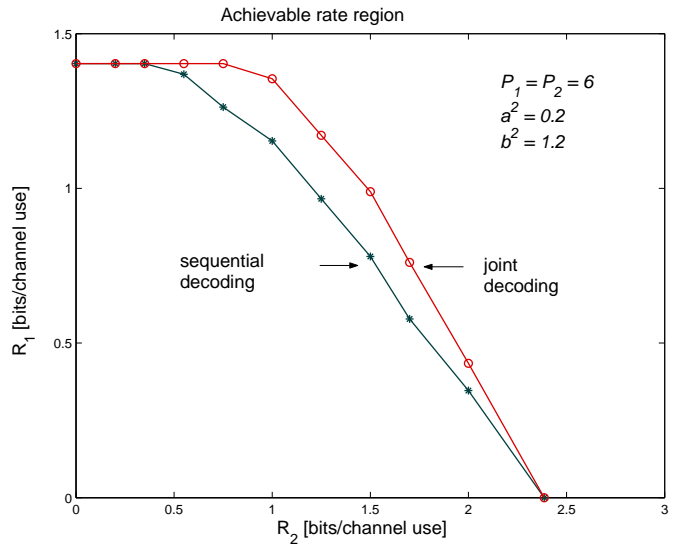,width=3.5in} \caption[Interference
channel with unidirectional cooperation.]{Comparison of achievable rates with joint and sequential decoding.}
      \label{SeqDecodingAug15}
\end{figure}

Fig.~\ref{SeqDecodingAug15} shows the performance of the two decoding schemes given the channel gain values $a$ and $b$ for which the performance differs significantly.
%%%%%%%%%%%%%%%%%%%%%%%%%%%%%%%%%%%%%%%%%%%%%%%%%%%%%%%%%%%%%%%%%%%%%%%%%%%%%%%%%%%%%%%%%%%%%%%%%%%%%%
%%%%%%%%%%%%%%%%%%%%%%%%%%%%%%%%%%%%%%%%%%%%%%%%%%%%%%%%%%%%%%%%%%%%%%%%%%%%%%%%%%%%%%%%%%%%%%%%%%%%%
%%%%%%%%%%%%%%%%%%%%%%%%%%%%%%%%%%%%%%%%
%\section{Strong Interference}
%%%%%%%%%%%%%%%%%%%%%%%%%%%%%%%%%%%%%%%%
\section{Conclusions and Future Work}
%%%%%%%%%%%%%%%%%%%%%%%%%%%%%%%%%%%%%%%%%
We have developed an encoding strategy for the interference channel with one
 cooperating encoder that generalizes previously proposed
encoding strategies. We evaluated its performance and compared it to
the performance of other schemes, focusing on the Gaussian channel.
A comparison with \cite{DevroyeSubmitted} would be an interesting next step. It is unclear whether our strategy generalizes the scheme in \cite{DevroyeSubmitted}, or whether a combination of the two techniques would achieve higher rates.
We also compared the proposed scheme to the outer bound that we developed for the strong interference regime.
 We further developed a new outer bound that extends the Nair-El
Gamal broadcast outer bound. Evaluating this bound for specific channels such as Gaussian may prove useful.

The cognitive radio channel shares some characteristics of both interference channels and broadcast channels. Combining encoding strategies developed for either of the two channel models therefore seems a natural approach. However, the optimality of a particular encoding scheme seems to be in part dictated by the channel conditions: for the Gaussian channel in which decoder $2$ experiences weak interference, dirty-paper coding is capacity achieving. On the other hand, strong interference conditions may allow the cognitive receiver  to decode the message not intended for him and therefore DPC against that message is not needed; superposition coding and rate-splitting achieve capacity. An even simpler scheme suffices when both receivers experience strong interference and can both decode the two messages. Neither DPC nor rate-splitting is needed; superposition coding is capacity-achieving.
The encoding scheme presented in this paper is a combination of rate-splitting, GP binning and superposition coding.
We believe that this general encoding scheme may be capacity-achieving for certain special cases related to the channel or specific encoding/decoding constraints. Finding such special cases is a topic of ongoing investigation
 %and we hope that it will lead to capacity results for special cases or more demanding decoding %constraints, not yet recognized in this paper.
%%%%%%%%%%%%%%%%%%%%%%%%%%%%%%%%%%%%%%%%%%%
%%%%%%%%%%%%%%%%%%%%%%%%%%%%%%%%%%%%%%%%
\section*{Appendix A: Proof of Theorem~\thmref{jointdecoding}} \label{A}
%%%%%%%%%%%%%%%%%%%%%%%%%%%%%%%%%%%%%%%%
%%%%%%%%%%%%%%%%%%%%%%%%%%%%%%%%%%%%%%%%%%%%%%%%%%%%%%%%%%%%%%%%%%
\begin{proof} {\it (Theorem~\thmref{jointdecoding})}
%%%%%%%%%%%%%%%%%%%%%%%%%%%%%%%%%%%%%%%%%%%%%%%%%%%%%%%%%%%%%%%%%%
%%%%%%%%%%%%%%%%%%%%%%%%%%%%%%%%%%%%%%%%%%%%%%%%
\begin{table*}
\begin{center}
\begin{minipage}{31.75pc}
\begin{center}
\begin{tabular*}{31.75pc}{|c|c|c|}
\hline
 & Error event & Arbitrarily small positive error probability if \\\hline
 $E_1$ & $(\hat w_c \neq 1, \hat w_{1a}=1)$ & $R_c + R'_c \le I(U_{1c}, U_{1a}; Y_1)$\\\hline
  $E_2$ & $(\hat w_c = 1, \hat w_{1a} \neq 1)$ & $R_{1a} + R'_{1a} \le I(U_{1a}; Y_1|U_{1c})$\\\hline
  $E_3$ & $(\hat w_c \neq 1, \hat w_{1a} \neq 1)$ & $R_c + R'_c + R_{1a} + R'_{1a} \le I(U_{1c}, U_{1a}; Y_1)$\\\hline
$E'_1$ & $(\hat w'_{2a} \neq 1, \hat w'_{2b}=1, \hat w'_c=1)$ &
$R_{2a}\le I(X_{2a}, X_{2b}; Y_2, U_{1c})$\\\hline
    $E'_2$ & $(\hat w'_{2a} \neq 1, \hat w'_{2b} \neq 1, \hat w'_c=1)$ & $R_{2a} + R_{2b} \le I(X_{2a}, X_{2b}; Y_2, U_{1c})$\\\hline
     $E'_3$ & $(\hat w'_{2a} \neq 1, \hat w'_{2b}=1, \hat w'_c \neq 1)$ & $R_{2a} + R_c + R'_c \le I(X_{2a}, X_{2b}, U_{1c}; Y_2) + I(U_{1c};X_{2a}, X_{2b})$\\\hline
      $E'_4$ & $(\hat w'_{2a} \neq 1, \hat w'_{2b} \neq 1, \hat w'_c \neq 1)$ & $R_{2a}+R_{2b} + R_c +R'_c \le I(X_{2a}, X_{2b}, U_{1c}; Y_2) + I(U_{1c};X_{2a}, X_{2b})$\\\hline
       $E'_5$ & $(\hat w'_{2a} = 1, \hat w'_{2b} \neq 1, \hat w'_c=1)$ & $R_{2b} \le I(X_{2b}; Y_2, U_{1c}|X_{2a})$\\\hline
        $E'_6$ & $(\hat w'_{2a} = 1, \hat w'_{2b} \neq 1, \hat w'_c \neq 1)$ & $R_{2b}+R_c + R'_c \le I(X_{2b}, U_{1c}; Y_2|X_{2a})+ I(U_{1c};X_{2a}, X_{2b})$\\\hline
\end{tabular*}
\end{center}
\end{minipage}
\end{center}
\medskip
\caption{Error events in joint decoding and corresponding rate
bounds.} \label{tab:errorevents}
\end{table*}
%%%%%%%%%%%%%%%%%%%%%%%%%%%%%%%%%%%%%%%%%%%%%%%%%%
{\bf Code construction:}  Ignore $Q$. Choose a distribution
$p(x_{2a},x_{2b},u_{1c},u_{1a},x_1,x_2)$.
\begin{itemize}
\item Split the rates as in \eqnref{ratesplitting1}-\eqnref{ratesplitting2}.
\item Generate $2^{NR_{2a}}$ codewords $x^N_{2a}(w_{2a})$, $w_{2a} = 1, \ldots , 2^{NR_{2a}}$, by choosing $x_{2a,n}(w_{2a})$ independently according to $P_{X_{2a}}(\cdot)$.
\item For each $w_{2a}$: Generate $2^{NR_{2b}}$ codewords $x^N_{2b}(w_{2a},w_{2b})$
using $\prod_{n=1}^N P_{X_{2b}|X_{2a}}(\cdot|x_{2a,n}(w_{2a}))$,
$w_{2b} = 1, \ldots , 2^{NR_{2b}}$.
\item For each pair $(w_{2a},w_{2b}):$ Generate
$x^N_2(w_{2a},w_{2b})$ where $x_2$ is a deterministic function of $(x_{2a},x_{2b})$.
\item Generate $2^{N(R_{1c}+R_{1c'})}$ codewords $u^N_{1c}(w_{c}, b_{c})$, $w_{c}=1, \ldots ,
2^{N{R_{1c}}}$, $b_c=1, \ldots , 2^{N{R_{1c'}}}$ using
$P_{U_{1c}}(\cdot).$
\item For each $u^N_{1c}(w_c,b_c)$:
Generate $2^{N(R_{1a}+R_{1a}')}$ codewords $u^N_{1a}( w_c,b_c,
w_{1a},b_{1a})$, $w_{1a}=1, \ldots , 2^{N{R_{1a}}}$, $b_{1a}=1,
\ldots , 2^{N{R_{1a}'}}$ using $\prod_{n=1}^N
P_{U_{1a}|U_{1c}}(\cdot|u_{1c,n}(w_c,b_c))$.
\item For $(w_1,w_2):$ Generate $x^N_1(w_{2a},w_{2b},w_c, b_c,w_{1a},b_{1a})$ where $x_1$ is a deterministic function of $(x_{2a},x_{2b},u_{1c},u_{1a},x_2)$.
\end{itemize}
Following the proof in \cite[Appendix D]{WillemsCribbing}, it can be shown that
it is enough to choose respective $x_2$ and $x_1$ to be
deterministic functions of $(x_{2a},x_{2b})$ and
$(x_{2a},x_{2b},u_{1c},u_{1a},x_2)$.

%%%%%%%%%%%%%%%%%%%%%%%%%%%%%%%%%%%%%%%%%%%%%%%%%%%
{\bf Encoders:} Encoder $1$:
\begin{enumerate}
\item Split the $NR_1$ bits $w_1$ into $NR_{1a}$ bits $w_{1a}$ and  $NR_c$ bits $w_c$.
Similarly, split the $NR_2$ bits $w_2$ into $NR_{2a}$ bits $w_{2a}$
and $NR_{2b}$ bits $w_{2b}$. We write this as
\begin{displaymath}
w_1 = (w_{1a}, w_c), \qquad w_2 = (w_{2a}, w_{2b}).
\end{displaymath}
\item  Try to find a bin index $b_c$ so that
$(u^N_{1c}(w_c,b_c),x^N_{2a}(w_{2a}),x^N_{2b}(w_{2a},w_{2b}) )\in
T_{\epsilon}(P_{U_{1c} X_{2a} X_{2b}})$ where $T_{\epsilon}(P_{XY})$
denotes jointly $\epsilon$-typical set with respect to $P_{XY}$, see
\cite[Sect.$8.6$]{Cover}. If  no such $b_c$ is found, choose
$b_c=1.$

\item For each $(w_c,b_c)$:
Try to find a bin index $b_{1a}$ such that $(\uv_{1a}(w_c,b_c,
w_{1a},b_{1a}), \uv_{1c}(w_c,b_c)), \xv_{2a}(w_{2a}), $ $
\xv_{2b}(w_{2a},w_{2b}) \in T_{\epsilon}(P_{U_{1a} U_{1c}X_{2a}
X_{2b}})$. If  cannot, choose $b_{1a}=1.$
\item Transmit $\xv_1(w_{2a},w_{2b},w_c, b_c,w_{1a},b_{1a})$.
\end{enumerate}

 Encoder $2$:
 Transmit $\xv_{2}(w_{2a},w_{2b})$.
%%%%%%%%%%%%%%%%%%%%%%%%%%%%%%%%%%%%%%%%%%%%%%%%%%%%%%%%%%%%%%%%%%%%%%%%%%%%%%%%%%%%%%%

{\bf Decoders:} Decoder $1$: Given $\yv_1$, choose $(\hat w_c, \hat
b_c, \hat w_{1a}, \hat b_{1a})$  if $( \uv_{1c}(\hat w_c, \hat b_c),
\uv_{1a}(\hat w_c, \hat b_c, \hat w_{1a}, \hat b_{1a}),  \yv_1) \in
T_{\epsilon}(P_{U_{1c}U_{1a}Y_1}).$
 If there is more than one such a quadruple, choose one.
 If there is no such quadruple, choose $(1,1,1,1)$

 Decoder $2$:
 Given $\yv_2$, choose $(\hat w_{2a}', \hat w_c', \hat b_c',\hat w_{2b}')$  if $( \xv_{2a}(\hat w_{2a}'), \uv_{1c}(\hat w_c', \hat b_c'), \xv_{2b}(\hat w_{2a}', \hat w_{2b}'),
 \yv_2)$ $\in
T_{\epsilon}(P_{X_{2a}U_{1c}X_{2b}Y_2}).$
 If there is more than one such a quadruple, choose one.
  If there is no such quadruple, choose $(1,1,1,1)$.

%%%%%%%%%%%%%%%%%%%%%%%%%%%%%%%%%%%%%%%%%%%%%%%%%%%%
{\bf Analysis:} Assume $(w_{1a}, w_c, w_{2a}, w_{2b})=(1,1,1,1)$ was
sent.
%%%%%%%%%%%%%%%%%%%%%%%%%%%%%%%%%%%%%%%%%%%%%%%%%%%%
Encoder error occurs if
\begin{enumerate}
\item Encoder $1$ cannot find a bin index $b_c$ such that
$(u^N_{1c}(1,b_c),x^N_{2a}(1),x^N_{2b}(1,1) )\in
T_{\epsilon}(P_{U_{1c} X_{2a} X_{2b}})$ which  happens with
probability
\begin{align}
&P_{e,enc1}^{(1)} = P [ \bigcap_{b_c=1}^{2^{NR_c'}}
(U^N_{1c}(1,b_c),x^N_{2a}(1),x^N_{2b}(1,1) )  \nonumber \\
& \qquad \qquad \quad  \notin T_{\epsilon}(P_{U_{1c} X_{2a} X_{2b}})
] \nonumber \\
&=\left( 1- P \left[ (U^N_{1c}(1,b_c),x^N_{2a}(1),x^N_{2b}(1,1) )
 \right. \right. \nonumber \\
& \qquad \left. \left. \in
T_{\epsilon}(P_{U_{1c} X_{2a} X_{2b}}) \right] \right)^{2^{NR_c'}} \nonumber \\
&\le \left( 1- (1-\epsilon)2^{-N[I(U_{1c};X_{2a},
X_{2b})+\delta]} \right)^{2^{NR_c'}} \nonumber \\
 &\le \exp \left( -
(1-\epsilon)2^{N[R'_c-I(U_{1c};X_{2a}, X_{2b})-\delta]} \right)
\eqnlabel{eqn9}
\end{align}
where the first inequality follows from \cite[Thm.$8.6.1$]{Cover}
and the second from $(1-x)^m \le \exp^{-mx}$. From \eqnref{eqn9},
the error probability $P_{e,enc1}^{(1)}$ can be made arbitrarily
small if
\begin{equation}
R'_c>I(U_{1c};X_{2a}, X_{2b})+\delta.
\end{equation}
\item After it has determined bin index $b_c$, say $b_c=1$, encoder $1$ cannot find a bin index $b_{1a}$ such that
$(u^N_{1a}(1,1, 1, b_{1a}), u^N_{1c}(1,1),x^N_{2a}(1),x^N_{2b}(1,1)
)\in T_{\epsilon}(P_{U_{1a} U_{1c} X_{2a} X_{2b}})$ which  happens
with probability
\begin{align}
&P_{e,enc1}^{(2)} = P [ \bigcap_{b_{1a}=1}^{2^{NR_{1a}'}}
(U^N_{1a}(1,1, 1, b_{1a}),u^N_{1c}(1,1),x^N_{2a}(1), \nonumber \\
& \qquad \qquad \quad  x^N_{2b}(1,1) ) \notin
T_{\epsilon}(P_{U_{1a}U_{1c} X_{2a}
X_{2b}})] \nonumber \\
&=\left( 1- P \left[ (U^N_{1a}(1,1,1,b_{1a}),
u^N_{1c}(1,1),x^N_{2a}(1),
 \right. \right. \nonumber \\
& \qquad \left. \left. x^N_{2b}(1,1) ) \in
T_{\epsilon}(P_{U_{1a}U_{1c} X_{2a} X_{2b}}) \right]
\right)^{2^{NR_{1a}'}}.  \eqnlabel{binsize2}
\end{align}
We have
\begin{align}
&P[(U^N_{1a}(1,1,1, b_{1a}), u^N_{1c}(1,1),x^N_{2a}(1), \nonumber
\\
&x^N_{2b}(1,1)\in T_{\epsilon}(P_{U_{1a} U_{1c} X_{2a}
X_{2b}})] \nonumber \\
&=\sum_{u_{1a}^N \in T_{\epsilon}(P_{U_{1a}U_{1c}X_{2a}X_{2b}}|u_{1c}^N x_{2a}^N x_{2b}^N)} P[u_{1a}^N| u_{1c}^N] \nonumber \\
&\ge (1-\epsilon)2^{-N(H(U_{1a}| U_{1c} X_{2a}X_{2b}
)-H(U_{1a}|U_{1c})+\delta)}   \nonumber \\
&=(1-\epsilon)2^{-N(I(U_{1a} ; X_{2a}X_{2b}|U_{1c})+\delta)}
\eqnlabel{important2}
\end{align}
where the first inequality follows from the fact that $U_{1a}^N$ was
generated according to $P_{U_{1a}|U_{1c}}$ (also directly from \cite[Handout $3.$]{KramerNotes}).

Employing \eqnref{important2}, we can bound \eqnref{binsize2}  as
\begin{align}
  &P_{e,enc1}^{(2)}\le \left( 1-
(1-\epsilon)2^{-N[I(U_{1a};X_{2a},
X_{2b}| U_{1c})+\delta]} \right)^{2^{NR_{1a}'}} \nonumber \\
 &\le \exp \left( -
(1-\epsilon)2^{N[R'_{1a}-I(U_{1a};X_{2a}, X_{2b}| U_{1c})-\delta]}
\right). \eqnlabel{eqn9r}
\end{align}
 We need
\begin{equation}
R_{1a}'>I(U_{1a};X_{2a}, X_{2b}| U_{1c})+\delta.
\end{equation}
\end{enumerate}

%%%%%%%%%%%%%%%%%%%%%%%%%%%%%%%%%%%%%%%%%%%%%%%%
{\bf Decoder errors:}
%%%%%%%%%%%%%%%%%%%%%%%%%%%%%%%%%%%%%%%%%%%%%%%%
 Possible error events at decoders are shown in the first column of Table~\ref{tab:errorevents}.
 We next derive the corresponding rate bounds given in the second column of the same table, which guarantee that the error
probability of each event can be made small as $N$ gets large. Bounds for
$E_1, E'_1, E'_3$ are loose. The rest of the rate expressions in
Table~\ref{tab:errorevents} yield \eqnref{jdR1a}-\eqnref{jdR2bRc}.

%%%%%%%%%%%%%%%%%%%%%%%%%%%%%%
% E1
%%%%%%%%%%%%%%%%%%%%%%%%%%%%%%
Consider the probability of event $E_1:$
\begin{align}
P[\hat w_c &\neq 1, \hat w_{1a}=1] \nonumber
 =\sum_{w_c=2}^{2^{NR_c}} \sum_{b_c=1}^{2^{NR'_c}}  P [\left( U_{1c}^N(w_c, b_c), \right. \\
&\left.  U_{1a}^N(w_c,b_c,1,1), Y^N_1 \right ) \in T_{\epsilon}(P_{
U_{1c} U_{1a} Y_1})]  \nonumber \\
& \le 2^{-N[I(U_{1c},U_{1a};Y_1) -(R_c+R'_c)-\delta]} \eqnlabel{E1}
\end{align}
%We have
%\begin{align}
%P&[(U_{1c}^N(w_c,b_c), U_{1a}^N(w_c,b_c,1,1),Y_1^N) \in
%T_{\epsilon}(P_{ U_{1c} U_{1a} Y_1})] \nonumber \\
%& \le 2^{-N[I(U_{1c}U_{1a};Y_1)-\delta]} \eqnlabel{E1P}
%\end{align}
by \cite[Thm.$8.6.1$]{Cover} and \cite[Handout $1$, Thm. $2$]{KramerNotes}.
From \eqnref{E1},  the arbitrarily small error
probability of $E_1$ requires
\begin{equation}
R_c + R'_c < I(U_{1c}, U_{1a};Y_1).
\end{equation}

%%%%%%%%%%%%%%%%%%%%%%%%%%%%%%
% E2
%%%%%%%%%%%%%%%%%%%%%%%%%%%%%%
Similarly, the probability of $E_2$ is
\begin{align}
P[\hat w_c &= 1, \hat w_{1a} \neq 1] \nonumber
 =\sum_{w_{1a}=2}^{2^{NR_{1a}}} \sum_{b_{1a}=1}^{2^{NR'_{1a}}}  P [\left( U_{1c}^N(1, 1), \right. \\
&\left.  U_{1a}^N(1,1,w_{1a},b_{1a}), Y^N_1 \right ) \in
T_{\epsilon}(P_{ U_{1c} U_{1a} Y_1})] \nonumber \\
& \le 2^{-N[I(U_{1a};Y_1|U_{1c})- (R_{1a}+R'_{1a})-\delta]}
\end{align}
where the inequality follows by \cite[Thm., Handout $3$]{KramerNotes}.
%We have
%\begin{align}
%P&[(U_{1c}^N(1,1), U_{1a}^N(1,1,w_{1a},b_{1a}),Y_1^N) \in
%T_{\epsilon}(P_{ U_{1c} U_{1a} Y_1})] \nonumber \\
%&= \sum_{(u_{1c}^N,u_{1a}^N,y_1^N) \in
%T_{\epsilon}}P[u_{1c}^Nu_{1a}^N] P[y_1^N|u_{1c}^N] \nonumber \\
%& \le 2^{-N[I(U_{1a};Y_1|U_{1c})-\delta]}
%\end{align}
We need
\begin{equation}
R_{1a} + R'_{1a} < I(U_{1a};Y_1|U_{1c}).
\end{equation}

%%%%%%%%%%%%%%%%%%%%%%%%%%%%%%
% E3
%%%%%%%%%%%%%%%%%%%%%%%%%%%%%%
The probability of $E_3$ is, similarly as in \eqnref{E1},
\begin{align}
P[\hat w_c & \neq 1, \hat w_{1a} \neq 1] \nonumber
 = \sum_{w_c=2}^{2^{NR_c}} \sum_{b_c=1}^{2^{NR'_c}} \sum_{w_{1a}=2}^{2^{NR_{1a}}} \sum_{b_{1a}=1}^{2^{NR'_{1a}}}  P [\left( U_{1c}^N(w_c, b_c), \right. \\
&\left.  U_{1a}^N(w_c,b_c,w_{1a},b_{1a}), Y^N_1 \right ) \in
T_{\epsilon}(P_{ U_{1c} U_{1a} Y_1})]. \nonumber \\
& \le 2^{-N[I(U_{1c},U_{1a};Y_1) -(R_c+R'_c+R_{1a} + R'_{1a}) -\delta]}
\end{align}
%As in \eqnref{E1P}, we have
%\begin{align}
%P&[(U_{1c}^N(w_c,b_c), U_{1a}^N(w_c,b_c,w_{1a},b_{1a}),Y_1^N) \in
%T_{\epsilon}(P_{ U_{1c} U_{1a} Y_1})] \nonumber \\
%& \le 2^{-N[I(U_{1c},U_{1a};Y_1)-\delta]}
%\end{align}
requiring
\begin{equation}
R_c+R'_c+R_{1a} + R'_{1a} < I(U_{1c},U_{1a};Y_1).
\end{equation}

%%%%%%%%%%%%%%%%%%%%%%%%%%%%%%%%%%%%%%%
% E1'
%%%%%%%%%%%%%%%%%%%%%%%%%%%%%%%%%%%%%%%
We next consider the error events at decoder $2$. For $E'_1$
\begin{align}
P[&\hat w'_{2a} \neq 1, \hat w'_{2b}=1, \hat w'_c = 1] \nonumber
 =\sum_{w_{2a}=2}^{2^{NR_{2a}}}  P [\left( U_{1c}^N(1, 1), \right. \\
&\left.  X_{2a}^N(w_{2a}), X_{2b}^N(w_{2a},1), Y^N_2 \right ) \in
T_{\epsilon}(P_{ U_{1c} X_{2a} X_{2b} Y_2})]. \label{E1prim}
\end{align}
We have
\begin{align}
P[(U_{1c}^N(1,1), &X_{2a}^N(w_{2a}), X_{2b}^N(w_{2a}, 1 ),Y_2^N) \nonumber \\
& \in
T_{\epsilon}(P_{ U_{1c} X_{2a} X_{2b} Y_2})] \nonumber \\
&= \sum_{(u_{1c}^N,x_{2a}^N,x_{2b}^N,y_2^N) \in
T_{\epsilon}}P[ x_{2a}^N, x_{2b}^N] P[y_2^N u_{1c}^N] \nonumber \\
& \le 2^{-N[I(X_{2a}, X_{2b};Y_2, U_{1c})-\delta]}. \label{E1Pprim}
\end{align}
From (\ref{E1prim}) and (\ref{E1Pprim}),
\begin{equation}
R_{2a} < I( X_{2a}, X_{2b};Y_2, U_c).
\end{equation}

%%%%%%%%%%%%%%%%%%%%%%%%%%%%%%%%%%%%%%%
% E2'
%%%%%%%%%%%%%%%%%%%%%%%%%%%%%%%%%%%%%%%
The probability of event $E'_2$ is
\begin{align}
P[&\hat w'_{2a} \neq 1, \hat w'_{2b} \neq 1, \hat w'_c = 1]
\nonumber
 =\sum_{w_{2a}=2}^{2^{NR_{2a}}} \sum_{w_{2b}=2}^{2^{NR_{2b}}} P [\left( U_{1c}^N(1, 1), \right. \\
&\left.  X_{2a}^N(w_{2a}), X_{2b}^N(w_{2a},w_{2b}), Y^N_2 \right ) \in
T_{\epsilon}(P_{ U_{1c} X_{2a} X_{2b} Y_2})]. \eqnlabel{E1prim}
\end{align}
Following  the same steps as in (\ref{E1Pprim}) and using
\eqnref{E1prim} it can be shown that the arbitrarily small error
probability of $E'_2$ requires
\begin{equation}
R_{2a}+ R_{2b} < I( X_{2a}, X_{2b};Y_2, U_{1c}).
\end{equation}
%%%%%%%%%%%%%%%%%%%%%%%%%%%%%%%%%%%%%%%%%%%%%%%%%%%%%%%%%%%%%%%%%%%%%%%%%%%%%%%%%%%%%%%%%%%%%%%%%
% E3'
%%%%%%%%%%%%%%%%%%%%%%%%%%%%%%%%%%%%%%%%%%%%%%%%%%%%%%%%%%%%%%%%%%%%%%%%%%%%%%%%%%%%%%%%%%%%%%%%%
 We next consider $E'_3$:
\begin{align}
P[&\hat w'_{2a} \neq 1, \hat w'_{2b} = 1, \hat w'_c \neq 1]
\nonumber \\
& =\sum_{w_{2a}=2}^{2^{NR_{2a}}}  \sum_{w_c=2}^{2^{NR_c}}  \sum_{b_c=1}^{2^{NR'_c}}
 P [\left( U_{1c}^N(w_c, b_c), X_{2a}^N(w_{2a}), \right. \\
&\left.   X_{2b}^N(w_{2a},1), Y^N_2 \right ) \in
T_{\epsilon}(P_{ U_{1c} X_{2a} X_{2b} Y_2})]. \eqnlabel{E3prim}
\end{align}
We have
\begin{align}
P[(U_{1c}^N&(w_c,b_c), X_{2a}^N(w_{2a}), X_{2b}^N(w_{2a}, 1 ),Y_2^N)
 \nonumber \\ & \qquad \qquad \in
T_{\epsilon}(P_{ U_{1c} X_{2a} X_{2b} Y_2})] \nonumber \\
&= \sum_{(u_{1c}^N,x_{2a}^N,x_{2b}^N,y_2^N) \in
T_{\epsilon}}P[ x_{2a}^N, x_{2b}^N] P[u_{1c}^N] P[y_2^N] \nonumber \\
& \le 2^{-N[I(X_{2a}, X_{2b}, U_{1c};Y_2)
+I(U_{1c};X_{2a},X_{2b})-\delta]}. \eqnlabel{E3Pprim}
\end{align}
From \eqnref{E3prim} and \eqnref{E3Pprim} it follows that
\begin{equation}
R_{2a}+ R_c + R'_c < I( X_{2a}, X_{2b}, U_{1c}; Y_2)
+I(U_{1c};X_{2a},X_{2b}).
\end{equation}
%%%%%%%%%%%%%%%%%%%%%%%%%%%%%%%%%%%%%%%%%%%%%%%%%%%%%%%%%%%%%%%%%%%%%%%%%%%%%%%%%%%%%%%%%%%%%%%%%
% E4'
%%%%%%%%%%%%%%%%%%%%%%%%%%%%%%%%%%%%%%%%%%%%%%%%%%%%%%%%%%%%%%%%%%%%%%%%%%%%%%%%%%%%%%%%%%%%%%%%%
For $E'_4$ we use the same approach as in
\eqnref{E3prim} and reuse \eqnref{E3Pprim} to obtain
\begin{equation}
R_{2a}+ R_{2b} + R_c + R'_c < I( X_{2a}, X_{2b}, U_{1c}; Y_2)
+I(U_{1c};X_{2a},X_{2b}).
\end{equation}

%%%%%%%%%%%%%%%%%%%%%%%%%%%%%%%%%%%%%%%%%%%%%%%%%%%%%%%%%%%%%%%%%%%%%%%%%%%%%%%%%%%%%%%%%%%%%%%%%
% E5'
%%%%%%%%%%%%%%%%%%%%%%%%%%%%%%%%%%%%%%%%%%%%%%%%%%%%%%%%%%%%%%%%%%%%%%%%%%%%%%%%%%%%%%%%%%%%%%%%%
We continue by considering error event $E_5'$:
\begin{align}
P[&\hat w'_{2a} = 1, \hat w'_{2b} \neq 1, \hat w'_c = 1] \nonumber
 =\sum_{w_{2b}=2}^{2^{NR_{2b}}}
 P [\left( U_{1c}^N(1,1), \right. \\
&\left.  X_{2a}^N(1), X_{2b}^N(1, w_{2b}), Y^N_2 \right ) \in
T_{\epsilon}(P_{ U_{1c} X_{2a} X_{2b} Y_2})]. \eqnlabel{E4prim}
\end{align}
We have
\begin{align}
P[(U_{1c}^N&(1,1), X_{2a}^N(1), X_{2b}^N(1,w_{2b} ),Y_2^N) \in
T_{\epsilon}(P_{ U_{1c} X_{2a}, X_{2b}, Y_2})] \nonumber \\
&= \sum_{(u_{1c}^N,x_{2a}^N,x_{2b}^N,y_2^N) \in
T_{\epsilon}}P[ x_{2a}^N, x_{2b}^N] P[u_{1c}^N,y_2^N|x_{2a}^N] \nonumber \\
& \le 2^{-N[I( X_{2b}, U_{1c};Y_2|X_{2a})-\delta]}. \eqnlabel{E4Pprim}
\end{align}
From \eqnref{E4prim} and \eqnref{E4Pprim} it follows that
\begin{equation}
R_{2b}< I( X_{2b}, U_{1c}; Y_2| X_{2a}).
\end{equation}

%%%%%%%%%%%%%%%%%%%%%%%%%%%%%%%%%%%%%%%%%%%%%%%%%%%%%%%%%%%%%%%%%%%%%%%%%%%%%%%%%%%%%%%%%%%%%%%%%
% E6'
%%%%%%%%%%%%%%%%%%%%%%%%%%%%%%%%%%%%%%%%%%%%%%%%%%%%%%%%%%%%%%%%%%%%%%%%%%%%%%%%%%%%%%%%%%%%%%%%%
For the error event $E'_6$ we have
\begin{align}
P[&\hat w'_{2a} = 1, \hat w'_{2b} \neq 1, \hat w'_c \neq 1]
\nonumber
 =\sum_{w_{2b}=2}^{2^{NR_{2b}}}  \sum_{w_c=2}^{2^{NR_c}} \sum_{b_c=1}^{2^{NR'_c}}
 P [\left( U_{1c}^N(w_c,b_c), \right. \\
&\left.  X_{2a}^N(1), X_{2b}^N(1, w_{2b}), Y^N_2 \right ) \in
T_{\epsilon}(P_{ U_{1c} X_{2a} X_{2b} Y_2})]. \eqnlabel{E6prim}
\end{align}
Again
\begin{align}
P[(U_{1c}^N&(w_c,b_c), X_{2a}^N(1), X_{2b}^N(1,w_{2b} ),Y_2^N) \nonumber \\
\nonumber & \qquad \qquad \in
T_{\epsilon}(P_{ U_{1c} X_{2a} X_{2b} Y_2})] \nonumber \\
&= \sum_{(u_{1c}^N,x_{2a}^N,x_{2b}^N,y_2^N) \in
T_{\epsilon}}P[ x_{2a}^N, x_{2b}^N] P[u_{1c}^N] P[y_2^N|x_{2a}^N] \nonumber \\
& \le 2^{-N[I( X_{2b}, U_{1c};Y_2|X_{2a})
+I(U_{1c};X_{2a},X_{2b})-\delta]}. \eqnlabel{E6Pprim}
\end{align}
From \eqnref{E6prim} and \eqnref{E6Pprim} it follows that
\begin{equation}
R_{2b}+R_c+R'_c< I( X_{2b}, U_{1c}; Y_2| X_{2a})+I(  U_{1c};
X_{2a},X_{2b}).
\end{equation}
%%%%%%%%%%%
\end{proof}
%%%%%%%%%%%%%%%%%%%%%%%%%%%%%%%%%%%%%%%%%%%%%%%%%%%%%%%%%%%%%%%%%%%%%%%%%%%%
\section*{Appendix B: Proof of Lemma~\lemmaref{GerhardLemma}} \label{B}
%%%%%%%%%%%%%%%%%%%%%%%%%%%%%%%%%%%%%%%%%%%%%%%%%%%%%%%%%%%%%%%%%%%%%%%%%%%%%
\begin{proof} {\it (Lemma~\lemmaref{GerhardLemma})}
For $I(S;Y,U) \le R_s \le H(S)$ use G-P coding \cite{GelfandPinsker80}.
The achieved rate is
\begin{equation}
R \le I(U;Y)-I(U;S).
\end{equation}
Note that $I(U;Y)-I(U;S) \le I(X;Y|S)$.

For $R_s < I(S;Y,U)$ proceed as follows.

{\bf Code construction:}
For every codeword $s^N(j), j=1, \ldots, 2^{NR_s}$ generate $2^{NR}$
codewords $u^N(w,j)$, $w=1, \ldots, 2^{NR}$ using $\prod_{n=1}^N
P_{U|S}(\cdot|s_n(j)).$

{\bf Encoder:}
Given $w$ and $s^N(j)$, choose $u^N(w,j)$ and transmit $x^N=f^N(u^N(w,j), s^N(j))$.

{\bf Decoder:} Given $y^N$, try to find $(w,j)$ such that $
(u^N(w,j),s^N(j),y^N) \in T_{\epsilon}(P_{SUY}).$

{\bf Analysis:} Suppose $w=1, j=1$ was sent. Error $\{\hat w \neq 1
\}$ occurs if $\{  \hat w \neq 1, {\hat j} \neq 1  \}$ or $\{  \hat
w \neq 1, \hat j = 1 \}$. The probability of error is
\begin{align}
P_e =&\sum_{j=2}^{2^{NR_s}}
\sum_{w=1}^{2^{NR}}P[(U^N(w,j),S^N(j),Y^N) \in
T_{\epsilon}(P_{SUY})] \nonumber \\
& + \sum_{w=2}^{2^{NR}} P[(U^N(w,1),s^N(1),Y^N) \in
T_{\epsilon}(P_{SUY})] \nonumber \\
&\le 2^{N(R+R_s-I(US;Y)+\delta_1)} +2^{N(R-I(U;Y|S)+\delta_2)} \eqnlabel{Pebinning}
\end{align}
where $\delta_1, \delta_2 \rightarrow 0$ as $N \rightarrow 0$.
From \eqnref{Pebinning} and $I(U;Y|S) = I(X;Y|S)$ it follows that
\begin{equation}
R \le \min\{ I(U,S;Y)-R_s, I(X;Y|S) \}. \eqnlabel{81}
\end{equation}
Note that we could have chosen $u^N=x^N$ in the superposition coding above, so that \eqnref{81} is
\begin{equation}
R \le \min\{ I(X,S;Y)-R_s, I(X;Y|S) \}. \eqnlabel{82}
\end{equation}
\end{proof}
%%%%%%%%%%%%%%%%%%%%%%%%%%%%%%%%%%%%%%%%%%%%%%%%%%%%%%%%%%%%%%%%%%%%%
\section*{Appendix C: Proof of Theorem~\thmref{generalouterbound}} \label{C}
%%%%%%%%%%%%%%%%%%%%%%%%%%%%%%%%%%%%%%%%%%%%%%%%%%%%%%%%%%%%%%%%%%%%%
%%%%%%%%%%%%%%%%%%%%%%%%%%%%%%%%
\begin{proof} {\it (Theorem~\thmref{generalouterbound})}
 Consider a code $(M_1,M_2,N,P_e)$ for the interference channel with one cooperating encoder.
%%%%%%%%%%%%%%%%%%%%%%%%%%%
% R1+R2
%%%%%%%%%%%%%%%%%%%%%%%%%%%
We first consider the bound  \eqnref{R1R2u2}. Fano's inequality
implies that for reliable communication we require
\begin{align}
&N(R_1+R_2)  \nonumber \\
& \le I(W_1;\Yv_1)+I(W_2;\Yv_2) \nonumber\\
& \le^{(a)} I(W_1;\Yv_1|W_2)+I(W_2;\Yv_2) \nonumber \\
& = \sum_{i=1}^N I(W_1;Y_1^i|W_2,Y_{2,i+1}^N)
    - I(W_1;Y_1^{i-1}|W_2,Y_{2,i}^N) \nonumber \\
& \qquad    + I(W_2;Y_{2,i}|Y_{2,i+1}^N) \nonumber \\
&= \sum_{i=1}^N I(W_1;Y_1^i|W_2,Y_{2,i+1}^N)
    - [I(W_1,Y_{2,i};Y_1^{i-1}|W_2,Y_{2,i+1}^N) \nonumber \\
& \qquad         - I(Y_{2,i};Y_1^{i-1}|W_2,Y_{2,i+1}^N)]
    + I(W_2;Y_{2,i}|Y_{2,i+1}^N) \nonumber \\
&=^{(b)} \sum_{i=1}^N I(W_1;Y_{1,i}|W_2,V_i)
    - I(Y_{2,i};Y_1^{i-1}|W_1,W_2,Y_{2,i+1}^N) \nonumber \\
& \qquad    + I(W_2,Y_1^{i-1};Y_{2,i}|Y_{2,i+1}^N) \nonumber  \\
& \le \sum_{i=1}^N I(W_1;Y_{1,i}|W_2,V_i)
    + I(W_2,V_i;Y_{2,i}) \eqnlabel{R1R2outer}
\end{align}
where $(a)$ follows from the independence of $W_1,W_2$;
 %\eqnref{R1R2a} follows from  \eqnref{FI2} and \eqnref{FI1c}
 in $(b)$, we let
 $Y_{t,i}^j=(Y_{t,i}, \ldots , Y_{t,j})$ and $V_i=[Y_1^{i-1},Y_{2,i+1}^N].$

%%%%%%%%%%%%%%%%%
% R2
%%%%%%%%%%%%%%%%%
We next consider the bound \eqnref{R2u2}. Fano's inequality implies
\begin{align}
NR_2
& \le I(W_2;\Yv_2)  \nonumber \\
& = \sum_{i=1}^N I(W_2; Y_{2,i}| Y_{2,i+1}^N) \nonumber \\
%& \le \sum_{i=1}^N I(W_2, Y_{2,i+1}^N ; Y_{2,i}) \nonumber \\
& \le \sum_{i=1}^N I(W_2,  Y_1^{i-1}, Y_{2,i+1}^N; Y_{2,i}) \nonumber \\
& = \sum_{i=1}^N I(W_2,V_i; Y_{2,i}). \eqnlabel{R2b}
\end{align}

%%%%%%%%%%%%%%%%%%
% R1+R2
%%%%%%%%%%%%%%%%%%

Note that for  \eqnref{R1R2outer}-\eqnref{R2b} we have used only the
independence of $W_1$ and $W_2$, and the non-negativity of mutual
information. The bounds \eqnref{R1u2} and \eqnref{R1R20u2} thus
follow by symmetry.

%%%%%%%%%%%%%%%%%%%
% DISTRIBUTION
%%%%%%%%%%%%%%%%%%%
 We introduce random variables $U_{1,i}=W_1$ and $U_{2,i}=W_2$ for all $i$, to get the bounds in the form \eqnref{R1u2}-\eqnref{R1R2u2}.
 Observe that $U_{1,i}$ and $U_{2,i}$ are independent.
Furthermore, due to unidirectional cooperation,
%the following is a
%Markov chain
%\begin{equation} \eqnlabel{MC}
%U_{1,i} \rightarrow (V_i,U_{2,i}) \rightarrow X_{2,i}.
%\end{equation}
%Therefore,
the joint probability distribution factors as in \eqnref{pdf}.
%\begin{align}
%&p(w,u_1,u_2,
%x_1,x_2)\\
%&=p(u_1)p(u_2)p(w|u_1,u_2)p(x_1|w,u_1)p(x_2|w,u_1,u_2,x_1).
%\nonumber
%\end{align}
%%%%%%%%%%%%%
% FDG
%%%%%%%%%%%%%
%Markov chain \eqnref{MC} was not obvious to me and hence I used the
%technique \cite{Kramer03} that determines conditional independence
%of random variables by considering {\it functional conditional
%graphs} (FDGs).
%Following the notation in \cite{Kramer03}, we let $\Xc=\{U_2\}= \{
%W_2 \}, \Zc=\{ W,U_1 \} = \{ W_1, \Yv_{1,i-1}, \Yv_2^{i+1}\}$,  $\Yc
%=\{ X_1 \}$ . For our problem, FDG $\Gc$ representing all random
%variables is shown in Figure~\ref{f:G}. We form the subgraph
%$\Gc_{\Xc \Yc \Zc}$ of $\Gc$, shown in Figure~\ref{f:GXYZ}, that has
%only vertices in $\Xc, \Yc,\Zc$ and vertices and edges encountered
%when moving backward from vertices in $\Xc, \Yc,\Zc$. The subgraph
%of $\Gc_{\Xc \Yc \Zc}$, $\Gc_{\Xc \Yc| \Zc}$ formed by removing
%edges coming out of $\Zc$ is shown in Figure~\ref{f:GXY}. We observe
%that $X_{1,i}$ is disconnected from the rest of the graph and thus
%independent from $ W_2$. Therefore, given $ (W,U_1) $, random
%variables $X_1$ and $U_2$ are independent and there is a Markov
%chain \eqnref{MC}.
%%%%%%%%%%%%%%
\end{proof}
%%%%%%%%%%%%%%%%%%%%%%%%%%%%%%%%%%%%%%%%%%%%%%%%%%%%%%%%%%%%%%%%%%%%%%%%%%%%%
\section*{Appendix D: Proof of Theorem~\thmref{stronginterferenceouterbound}} \label{D}
%%%%%%%%%%%%%%%%%%%%%%%%%%%%%%%%%%%%%%%%%%%%%%%%%%%%%%%%%%%%%%%%%%%%%%%%%%%%%
\begin{proof} {\it (Theorem~\thmref{stronginterferenceouterbound})}
The bound \eqnref{R1si} follows by standard methods. To prove
\eqnref{R1R2si}, consider \eqnref{R1R2u2} and
\begin{align}
   I(U_1 ; Y_1 | U_2,V)&
   \le I(U_1 ; Y_1,X_2 | U_2,V) \nonumber \\
  & =   I(U_1 ; Y_1 | U_2,V,X_2)  \nonumber \\
  & \le I(U_1,X_1 ; Y_1 | U_2,V,X_2) \nonumber \\
  & =   I(X_1 ; Y_1 | U_2,V,X_2) \nonumber \\
  & \le I(X_1 ; Y_2 | U_2,V,X_2) \eqnlabel{bound1}
\end{align}
where the second step follows by the Markov chain \eqnref{pdf}, and
the last step follows by \eqnref{decodeboth}. We similarly have
\begin{equation}
I(V,U_2 ; Y_2) \le I(U_2,V,X_2 ; Y_2). \eqnlabel{bound2}
\end{equation}
Combining inequalities \eqnref{R1R2u2}, \eqnref{bound1} and
\eqnref{bound2} gives \eqnref{R1R2si}.
\end{proof}

%%%%%%%%%%%%%%%%%%%%%%%%%%%%%%%%%%%%%%%%%%%%
%\bibliographystyle{C:/texmf/bibtex/bst/ieeetran/IEEEtran}
%\bibliographystyle{C:/downloads/IEEEstyles/IEEEtran}
\bibliographystyle{IEEEtran}
%\bibliography{referencesCOOP}
%old laptop
%\bibliography{C:/IVANA/PAPERS/referencesCOOP}
%newlaptop
%\bibliography{C:/Users/Ivana/PAPERS/referencesCOOP}
%%%%%%%%%%%%%%%%%%%%%%%%%%%%%%%%%%%%%%%%%%%%

\end{document}